\documentclass[iop,twocolappendix]{emulateapj}

\usepackage{epsfig}
\usepackage{amsmath,amssymb}
\usepackage{ulem,}
\usepackage{graphicx}
\usepackage{natbib}
\usepackage[backref,breaklinks,colorlinks,citecolor=blue]{hyperref}

\usepackage{comment}
\font\sevenrm=cmr7 scaled 1000
\newcommand{\HeII}{He{\sevenrm II}$\lambda$4686}
\newcommand{\Msun}{$M_\odot$}
\newcommand{\Swift}{\textsl{Swift}}
\newcommand{\gptf}{$g_\textrm{PTF}$}
\newcommand{\rptf}{$R_\textrm{PTF}$}
\newcommand{\RNum}[1]{\uppercase\expandafter{\romannumeral #1\relax}}

\begin{document}

\title{Sifting for Sapphires: Systematic Selection of Tidal Disruption Events in \lowercase{i}PTF}
\author{T. Hung\altaffilmark{1}, S. Gezari\altaffilmark{1,2}, S.B. Cenko\altaffilmark{2,3}, S. van Velzen\altaffilmark{2,4}, N. Blagorodnova\altaffilmark{5}, Lin Yan\altaffilmark{6,7}, S. R. Kulkarni\altaffilmark{5},
R. Lunnan\altaffilmark{8}, T. Kupfer\altaffilmark{5}, G. Leloudas\altaffilmark{9}, A. K. H. Kong\altaffilmark{10}, P. E. Nugent\altaffilmark{11,12}, C. Fremling\altaffilmark{5}, Russ R. Laher\altaffilmark{7}, F. J. Masci\altaffilmark{7}, Y. Cao\altaffilmark{13,14}, R. Roy\altaffilmark{8}, and T. Petrushevska\altaffilmark{15}}
\altaffiltext{1}{Department of Astronomy, University of Maryland, College Park, MD 20742, USA}
\altaffiltext{2}{Joint Space-Science Institute, University of Maryland, College Park, MD 20742, USA}
\altaffiltext{3}{NASA Goddard Space Flight Center, Mail Code 661, Greenbelt, MD 20771, USA}
\altaffiltext{4}{Department of Physics, New York University, NY 10003, USA}
\altaffiltext{5}{Department of Astronomy, California Institute of Technology, Pasadena, CA 91125, USA}
\altaffiltext{6}{Caltech Optical Observatories, Cahill Center for Astronomy and Astrophysics, California Institute of Technology, Pasadena, CA 91125, USA}
\altaffiltext{7}{Infrared Processing and Analysis Center, California Institute of Technology, Pasadena, CA 91125, USA}
\altaffiltext{8}{The Oskar Klein Centre \& Department of Astronomy, Stockholm University, AlbaNova, SE-106 91 Stockholm, Sweden}
\altaffiltext{9}{Dark Cosmology Centre, Niels Bohr Institute, University of Copenhagen, Juliane Maries vej 30, 2100 Copenhagen, Denmark}
\altaffiltext{10}{Institute of Astronomy, National Tsing Hua University, No. 101, Section 2, Kuang-Fu Road, Hsinchu, 30013, Taiwan}
\altaffiltext{11}{Department of Astronomy, University of California, Berkeley, CA 94720-3411, USA}
\altaffiltext{12}{Lawrence Berkeley National Laboratory, 1 Cyclotron Road, MS 50B-4206, Berkeley, CA 94720, USA}
\altaffiltext{13}{Department of Astronomy, University of Washington, Box 351580, U.W., Seattle, WA 98195-1580, USA}
\altaffiltext{14}{eScience Institute, University of Washington, Box 351570, U.W., Seattle, WA 98195-1580, USA}
\altaffiltext{15}{Oskar Klein Centre, Department of Physics, Stockholm University, SE 106 91 Stockholm, Sweden}
\keywords{accretion, accretion disks -- black hole physics -- galaxies: nuclei -- ultraviolet: general}

\begin{abstract}

We present results from a systematic selection of tidal disruption events
(TDEs) in a wide-area (4800~deg$^2$), $g+R$ band, Intermediate
Palomar Transient Factory (iPTF) experiment.
Our selection targets typical optically-selected TDEs: bright ($>$60\% flux increase) and blue transients residing in the center of red galaxies.
Using photometric selection criteria to down-select from a total of 493 nuclear transients to a sample of 26 sources, we then use follow-up UV imaging with the Neil Gehrels Swift Telescope, ground-based optical spectroscopy, and light curve fitting to classify them as 14 Type Ia supernovae (SNe Ia), 9
highly variable active galactic nuclei (AGNs), 2 confirmed TDEs, and 1
potential core-collapse supernova. We find it possible to filter AGNs by
employing a more stringent transient color cut ($g-r <$ $-$0.2 mag); further, UV imaging is the best discriminator for filtering SNe, since SNe Ia can appear as blue, optically, as TDEs in their early phases.  However, when UV-optical color is unavailable, higher precision astrometry can also effectively reduce SNe contamination in the optical. Our most stringent optical photometric selection criteria
yields a 4.5:1 contamination rate, allowing for a manageable number of TDE
candidates for complete spectroscopic follow-up and real-time classification
in the ZTF era. We measure a TDE per galaxy
rate of 1.7$^{+2.9}_{-1.3}$ $\times$10$^{-4}$ gal$^{-1}$ yr$^{-1}$ (90\% CL in Poisson statistics). This does not
account for TDEs outside our selection criteria, thus may not reflect the
total TDE population, which is yet to be fully mapped.

\end{abstract}

\section{Introduction}
A stellar tidal disruption event (TDE) refers to the phenomenon of a star being scattered into the Roche radius of a supermassive black hole. For a star approaching on a parabolic orbit, about half of the stellar debris remains bound to the black hole after the disruption. In the paradigm of TDEs,
the circularized bound debris will form an accretion disk and emit thermally \citep{Rees1988,1999ApJ...514..180U} as it accretes onto the black hole. Alternatively, some studies have argued that the energy dissipated during the circularization process itself may be the main powering source of the emission in TDEs \citep{2015ApJ...806..164P,2015ApJ...804...85S,2015ApJ...812L..39D,2017MNRAS.464.2816B}.

With the advent of a number of wide-field time domain surveys, about a dozen convincing optically bright TDEs have been discovered. Unlike the first few TDEs discovered in X-ray surveys (e.g. ROSAT) that have sparse light curves, the dense light curve sampling (once every few days) by ground-based optical surveys have enabled timely discoveries of these optical TDEs. The well-sampled optical light curves, together with the rapid follow-up observations across the electromagnetic spectrum, have provided rich data sets that allow us to study these TDEs in more detail.
Although the selection criteria of these TDEs differ among surveys, most of them follow a power-law decline that is loosely consistent with t$^{-5/3}$ \citep{Phinney1989} in their light curves, while maintaining constant color over time. Most of these optically detected TDEs have a small spread in peak luminosity (10$^{43.4-44.4}$ erg s$^{-1}$) and blackbody temperature \citep[(2--4)$\times$10$^{4}$ K;][]{2017ApJ...842...29H}.

The first attempt of a systematic search of optically bright TDEs was made by \cite{vanVelzen2011} using archival SDSS Stripe 82 imaging data, where they carefully examined the transients in the nuclei of galaxies (hereafter nuclear transients) and successfully recovered two TDE candidates. The search for TDEs usually excludes active host galaxies, because most AGNs exhibit variability and they outnumber TDEs by at least two orders of magnitude. Follow-up on these sources is cost inefficient, since a detailed spectroscopic study would be required to differentiate between AGN variability and TDE signatures. Nevertheless, there is no mechanism prohibiting stars from being disrupted by active black holes. In fact, two events, CSS100217 \citep{2011ApJ...735..106D} and the recently reported PS16dtm \citep{2017ApJ...843..106B}, have been claimed as TDEs occurring in narrow-line Seyfert 1 galaxies. Their properties, however, can be somewhat different from the TDEs observed in quiescent galaxies. For example, the light curve of PS16dtm exhibits a plateau instead of a monotonical decline following $t^{-5/3}$, even though its blackbody temperature remained constant at $T_{bb} \approx$~1.7$\times$10$^4$~K. The broad H$\alpha$ component ($\approx$ 4000~km~s$^{-1}$) in both CSS100217 and PS16dtm are considerably narrower than TDEs in inactive galaxies \citep[10$^4$~km~s$^{-1}$;][]{2017ApJ...842...29H} and they exclusively exhibit strong Fe {\sevenrm II} emission complexes. In addition, both events were radiating near Eddington luminosity, which may suggest they are accreting more efficiently than TDEs in inactive galaxies \citep{2017ApJ...843..106B}.

Finding TDEs can also be complicated by the detection of transients in galaxy nuclei. Nuclear transient events suffer from stronger systematic and statistical errors when subtracted from the core of a galaxy. This poses a problem because large surveys like the intermediate Palomar Transient Factory (iPTF) rely on machine learning algorithms to remove image subtraction artifacts. Therefore, nuclear transients are more easily missed in this process of machine identification. Using the \texttt{realbogus} pipeline employed by iPTF as an example, the completeness drops from 1 to 0.5 when the flux contrast, defined as the ratio between the transient and the underlying host galaxy surface brightness, decreases from $>$10 times to 0.6 times \citep{2017ApJS..230....4F}.

While the majority of the optically discovered TDEs reside in red early-type galaxies (see color definition in \autoref{subsec:step3(host color)}), they seem to appear disproportionately more in a rare subset --- quiescent Balmer-strong (E+A) galaxies --- whose star formation history indicates they are in a post-starburst stage \citep{Arcavi2014,2016ApJ...818L..21F,2016ApJ...825L..14S}. On a different study that employs a control group from SDSS, \cite{2017arXiv170701559L} found that TDE host galaxies have a higher nuclear stellar density that is not owing to a recent major merger. Although the rare E+A galaxies are reported to be over-represented by at least 35 times in the current TDE sample \citep{2016ApJ...818L..21F,2017arXiv170702986G}, it is not known whether TDEs preferentially occur in E+A host galaxies, or they occur more frequently in centrally overdense galaxies.

The ever-growing number of TDEs discovered in optical surveys have gradually alleviated the tension between the observed and the predicted TDE rates, where the observed rate is typically an order of magnitude lower than the theoretical value predicted from two-body relaxation of stars in realistic galaxies \citep[$\gtrsim$ 10$^{-4}$ gal$^{-1}$ yr$^{-1}$;][]{1999MNRAS.309..447M,2004ApJ...600..149W,2016MNRAS.455..859S}.
It is generally accepted that the lower observational TDE rate can be attributed to the fact that only the brighter TDEs in a steep TDE luminosity function are followed-up in flux-limited surveys \citep{2017arXiv170703458V}.
The fact that the luminosities of most optical TDEs seem to be drawn from a distribution with a small spread \citep{2017ApJ...842...29H} is also consistent with this scenario as long as the TDE luminosity is not a narrow step function. Since it is almost impossible to obtain a spectroscopically complete sample for most of the photometric surveys except for the All-Sky Automated Survey for Supernova \citep[ASASSN has a flux limit of $r \sim 17$ mag;][]{2014ApJ...788...48S}, at least some parts of this bias is introduced by human decisions on whether to trigger spectroscopic classification for a certain object. Therefore, a systematic strategy for selecting TDE candidates in real time for follow-up observations is critical for measuring TDE rates robustly.

Future surveys such as the Zwicky Transient Facility (ZTF) in 2018 and the Large Synoptic Survey Telescope (LSST) in the early 2020s are expected to boost the number of transients including TDEs by orders of magnitude. Maximizing the efficiency of follow-up resources will be even more challenging. In 2016, iPTF conducted its first experiment with simultaneous color observations. With a Cycle 12 \Swift{} Key Project (PI Gezari) and observing time on several optical spectrographs, we were able to perform a systematic follow-up of the TDE candidates that satisfy the selection of our nuclear transient pipeline. Two TDEs, iPTF16axa ($z$=0.108) and iPTF16fnl ($z$=0.016), were discovered during this 4-month experiment. In particular, iPTF16fnl is the faintest and fastest TDE discovered yet with a decay rate of $\sim$0.08 mag day$^{-1}$ \citep{2017ApJ...844...46B}.

In this paper, we report the performance of our pipeline, which is optimized for TDEs by searching for blue (\gptf{}$-$\rptf{}$<$0 mag) nuclear flares within inactive red host galaxies, and derive the TDE rate in iPTF. We first describe the design of the \gptf{}+\rptf{} survey in \autoref{sec:design} and the selection of the nuclear candidates for \Swift{} and spectroscopic follow-up in \autoref{sec:selection}. The photometric and spectroscopic observations are detailed in \autoref{sec:observations}.
The classification of the final sample along with the results from \Swift{} are presented in \autoref{sec:classification}. We discuss the performance of our pipeline and calculate the TDE rate in iPTF in \autoref{sec:discussion}. We conclude by outlining the prospects of TDE discoveries with ZTF in \autoref{sec:discussion}.

\section{Rolling \lowercase{g}+R Survey Design}
\label{sec:design}
iPTF conducted a rolling $g+R$ experiment from UT 2016 May 28 to 2016 Dec 7, observing a total of 81 nights. The Palomar 48-inch (P48) telescope uses SDSS-$g'$ and Mould-$R$ filters (hereafter $g_\textrm{PTF}$ and $R_\textrm{PTF}$) to carry out the survey. The experiment was suspended between 2016 Jun 16 and 2016 Aug 20 when the telescope time was entirely dedicated to the H$\alpha$ survey and the Galactic plane survey.

The experiment took place in two stages. In the later period from 2016 Aug 20 to Nov 10, it was strictly controlled using a rolling strategy that was implemented as follows. We divided the 270 fields in total into three groups of 90 fields. At any given time, one of the groups was observed using a 1-day cadence, while the remaining 180 fields were observed using a 3-day cadence (i.e. 60 each night). As such, the total number of fields observed each night was 150. The 1-day cadence group was rotated so that every field would cycle between 1-day and 3-day cadences. This rotation took place regularly every lunation. Missed nights due to weather would become gaps in the light curve.
To obtain the color, each field was observed with one \gptf{} and one \rptf{} filter typically separated by $\sim$0.5~hr. This experiment was called the \textit{Color Me Intrigued} experiment when it was proposed \citep[see][for more details]{2017ApJ...848...59M}.

In the earlier stage from 2016 May 28 to Jun 16, fewer fields (58 fields on average) were targeted in a night. Each field was observed three times in \gptf{} and \rptf{} filters per night with two observations in the same filter. All of the targeted fields were scheduled to have a 4-day cadence. This change in cadence from the latter rolling cadence strategy should have little impact on the outcome of our TDE search, since tidal disruption flares can last for months. Both TDEs iPTF16axa and iPTF16fnl were discovered in the 3-d cadence fields.

\begin{figure*}
\centering
\includegraphics[width=7.0in, angle=0]{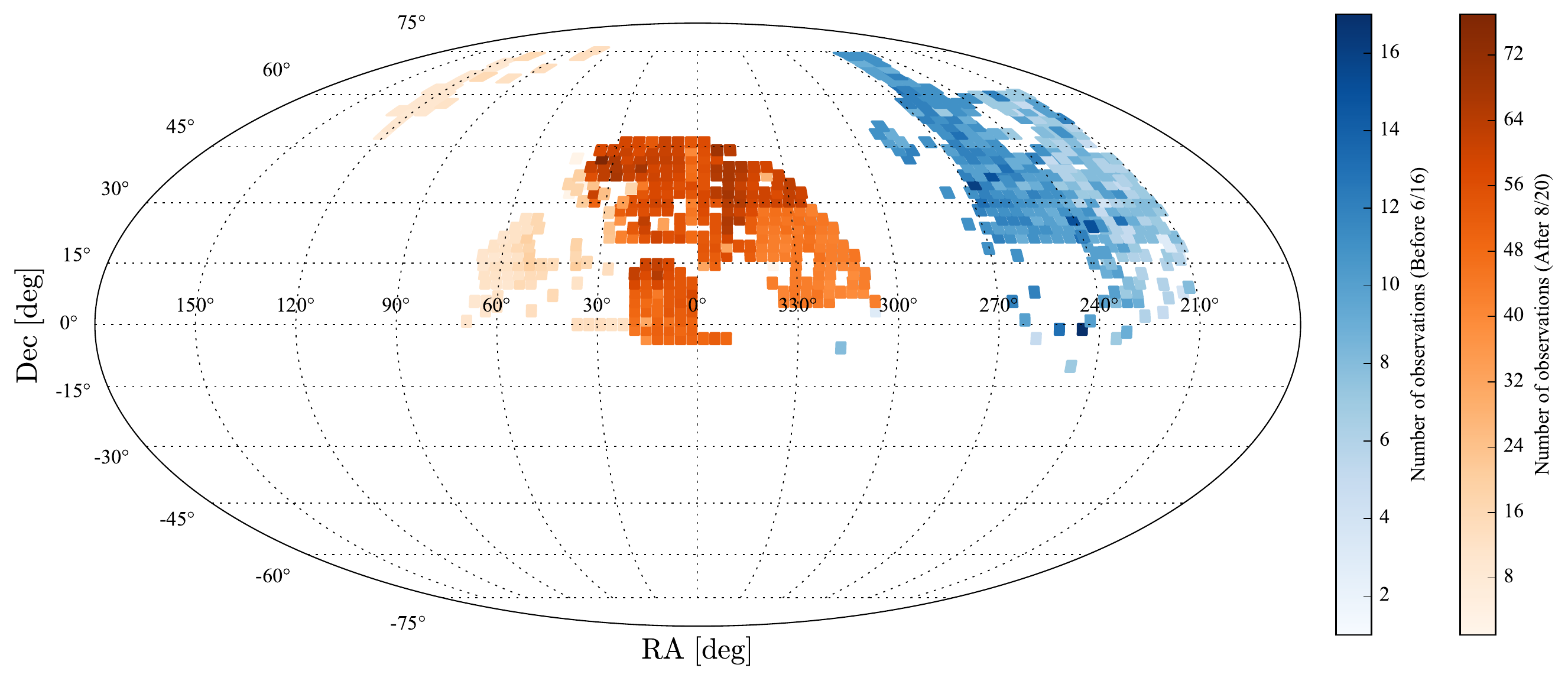}
\caption{Sky coverage of the iPTF color experiment in Equatorial coordinates. The plot is broken down by the stage of the survey (before and after the summer hiatus). Each tile represents a unique pointing, and the color intensity reflects the number of observations in that field. Around 77\% of the fields were also observed by SDSS.}
\label{fig:skymap}
\end{figure*}

\begin{figure*}
\centering
\includegraphics[width=7.0in, angle=0]{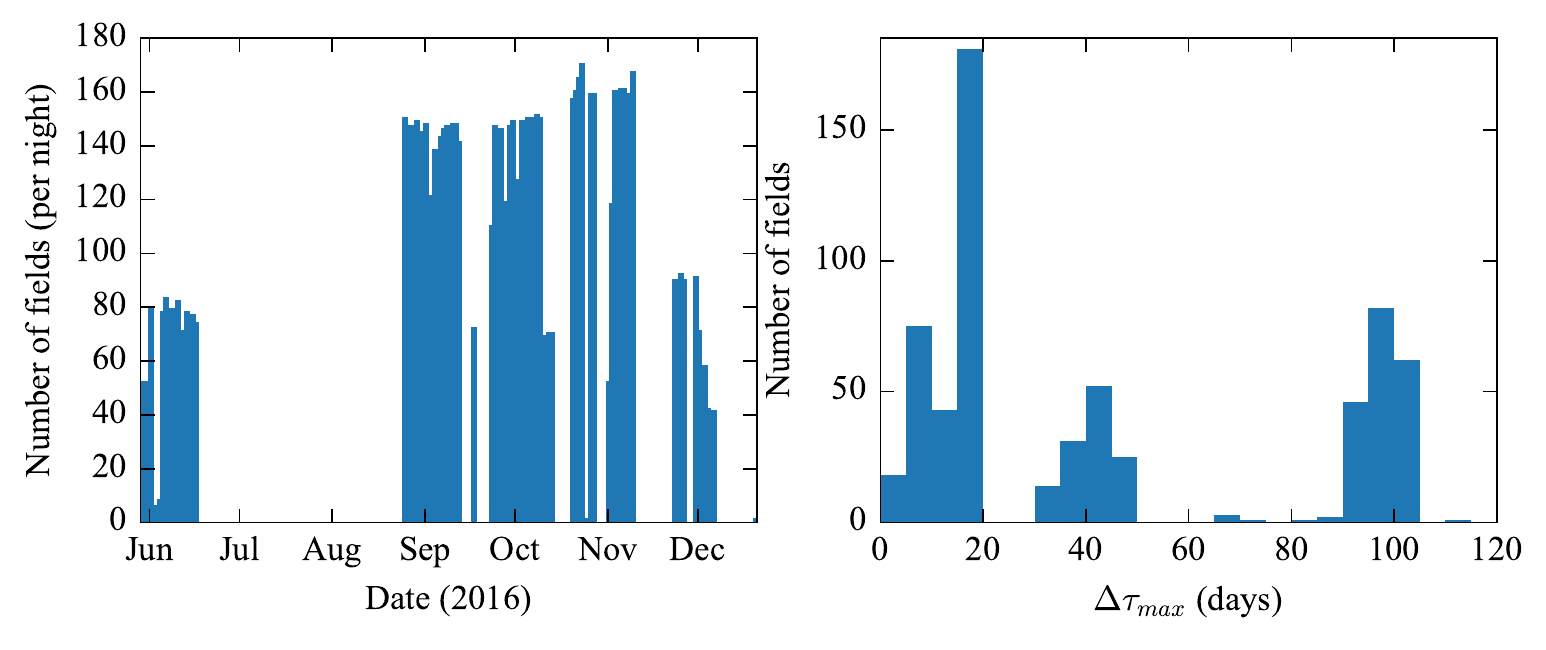}
\caption{Left: number of unique fields observed per night during the \gptf{}+\rptf{} survey. Fewer fields were observed before Aug 2016 because each field was observed three times (usually 2 in \gptf{} and 1 \rptf{}) per night instead of two. Right: distribution of the longest baseline of each field in the \gptf{}+\rptf{} experiment.}
\label{fig:grobs}
\end{figure*}

\begin{figure}
\centering
\includegraphics[width=3.5in, angle=0]{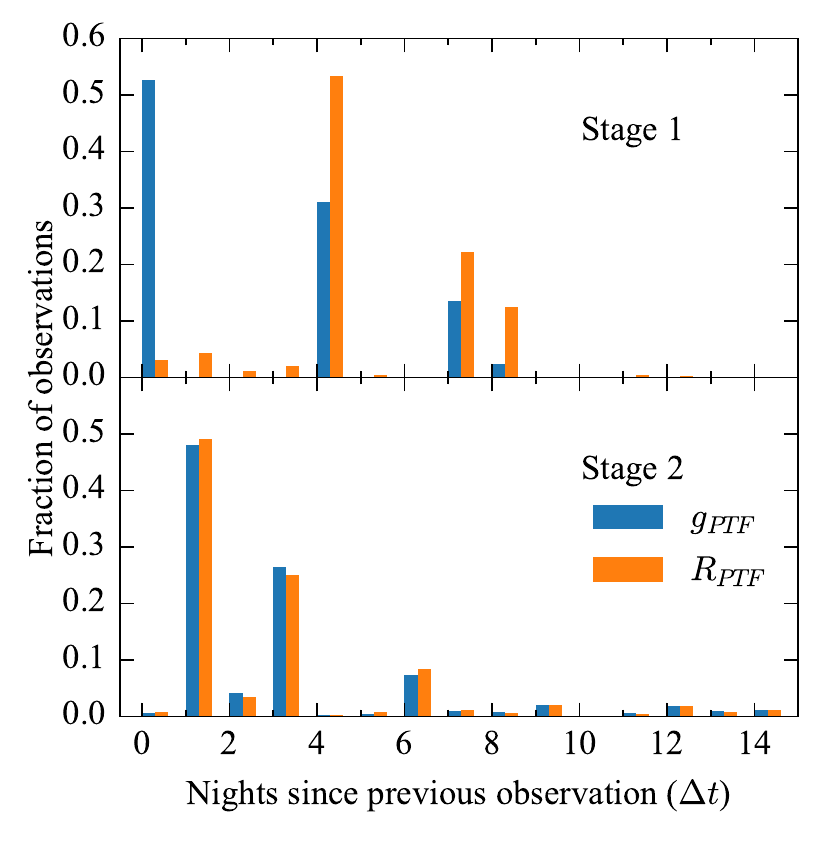}
\caption{Cadence of the \gptf{}+\rptf{} experiment in two stages. Plots show the fraction of $\Delta t$ between any two adjacent epochs for all the unique fields observed during this period in \gptf{} (blue) and \rptf{} (orange) bands. The left bin edges of \gptf{} band are aligned with the tick marks, with \rptf{} offset to the right for clarity. Intra-night observations ($\Delta t$=0) are dominated by the observations made before August (upper panel), where two observations in the same filter were used to reject moving objects. The designed cadence of the later stage (bottom panel) is a rotating $\Delta t$ of 1 and 3 days. The cadence for the later stage also clusters around 6 days due to weather loss.}
\label{fig:grcadence}
\end{figure}

We present the sky coverage of the $g+R$ experiment in \autoref{fig:skymap}, with color bars indicating the number of exposures. The fields were selected to be overhead and have low airmass at the Palomar observatory during the observing period.
We show the number of fields observed per night throughout the operation in the left panel of \autoref{fig:grobs}. Given the 7.26 deg$^2$ field of view of the telescope, this corresponds to an average areal coverage of 771 deg$^2$ per night. The right panel of \autoref{fig:grobs} shows the distribution of the longest baseline in each field excluding the summer gap. In \autoref{fig:grcadence}, the cadence distribution of the experiment is split by stages and by filters. In the early stage shown in the upper panel of \autoref{fig:grcadence}, most of the targeted fields were observed twice in \gptf{} band per night and have a 4-day cadence. The later stage shown in the lower panel peaks at 1, 3, 6 days as intended by the experiment strategy. In summary, The experiment monitored 660 unique fields, which corresponds to a total of 4792 deg$^2$ of the sky.

\section{Selection of candidates}
\label{sec:selection}
During the \gptf{}+\rptf{} experiment, the survey data was processed by the real-time image subtraction pipelines at the Infrared Processing and Analysis Center (IPAC) \citep{2017PASP..129a4002M} and the National Energy Research Scientific Computing Center (NERSC) \citep{2016PASP..128k4502C}. We rely on machine learning classification --- the probability based real-bogus (RB) score --- to determine the likelihood of the detection being a real astrophysical transient, as opposed to a false detection caused by image subtraction artifacts \citep{2017PASP..129a4002M}. Sources that had a S/N $>$ 5, a high RB score ($>$0.6), and at least 2 co-spatial detections within one night that were not cataloged as a variable star or a quasar, were visually inspected by a human. Since bad subtractions normally have an irregular PSF, they can be readily recognized by eye. This process, called scanning, took place on a daily basis. Only sources approved by the scanners were saved to the database for further study.

\begin{figure*}
\centering
\includegraphics[width=7.0in, angle=0]{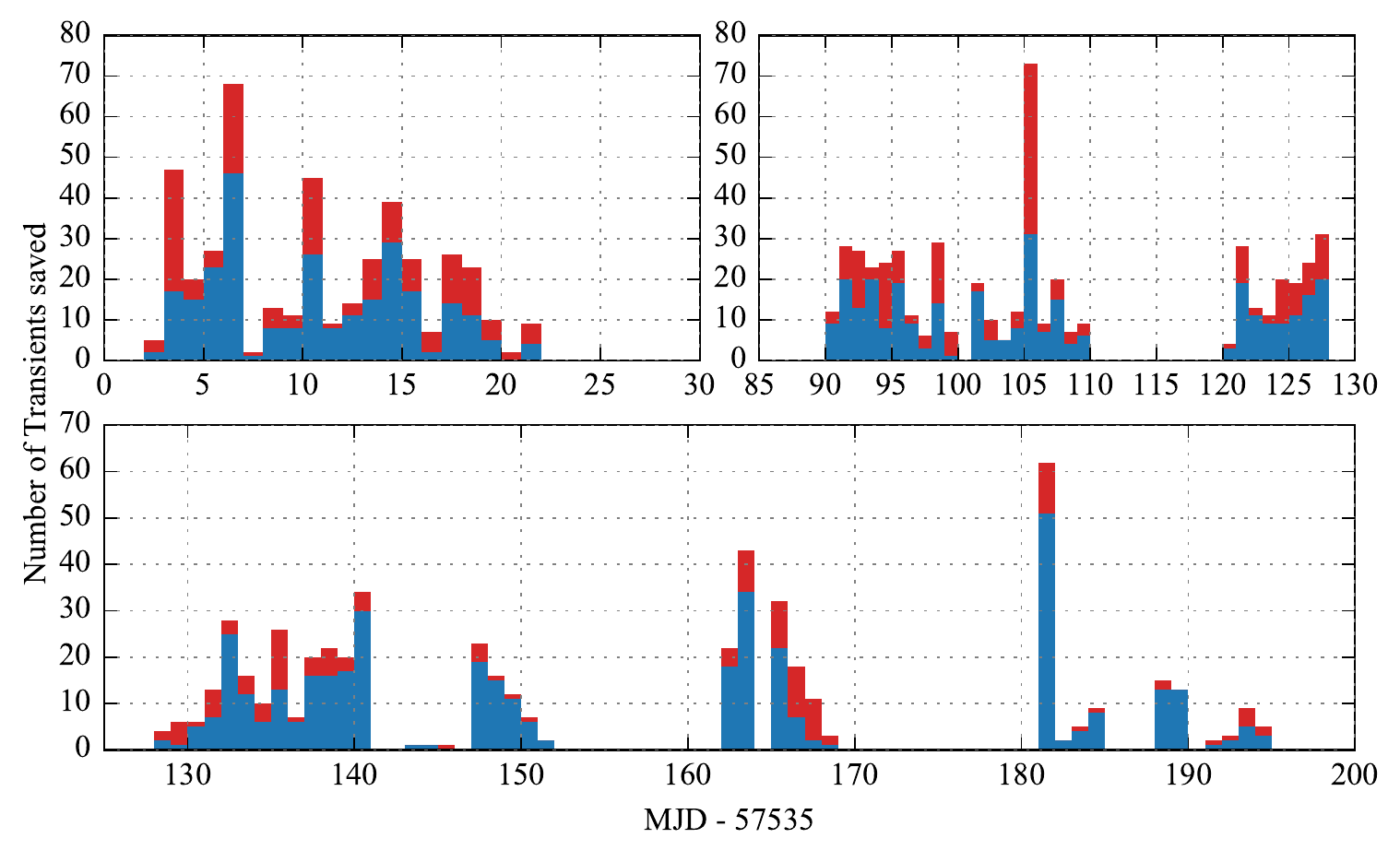}
\caption{Total number of nuclear (red) on top of non-nuclear (blue) transient events saved during the period of iPTF \gptf{}+\rptf{} experiment. Nuclear transients account for one-third of all the transients.}
\label{fig:transient}
\end{figure*}

We show the stacked histogram of the number of transients saved during this experiment in \autoref{fig:transient}, where the nuclear and non-nuclear transients are shown in red and blue, respectively. A total of 1464 transients were saved in this period, averaging 18 transients per night.

In transient surveys, the rate of SNe Ia \citep[2.7$\times$10$^{-5}$ yr$^{-1}$Mpc$^{-3}$;][]{Dilday2010} is on the order of 100 times higher than TDEs. Type 1 AGNs, which are known to be variable in the optical on timescales up to several years \citep{2007AJ....134.2236S}, have a space density of $\approx$ 7$\times$10$^{-6}$~Mpc$^{-3}$ \citep{2005AJ....129.1783H} that also outnumber TDEs by roughly two orders of magnitude. These two types of events are the main interlopers in the process of selecting TDE candidates. Because of their disproportionately higher occurrences, our selection criteria are focused on removing these interlopers, which we describe as follows.

\begin{figure}
\centering
\includegraphics[width=3.5in, angle=0]{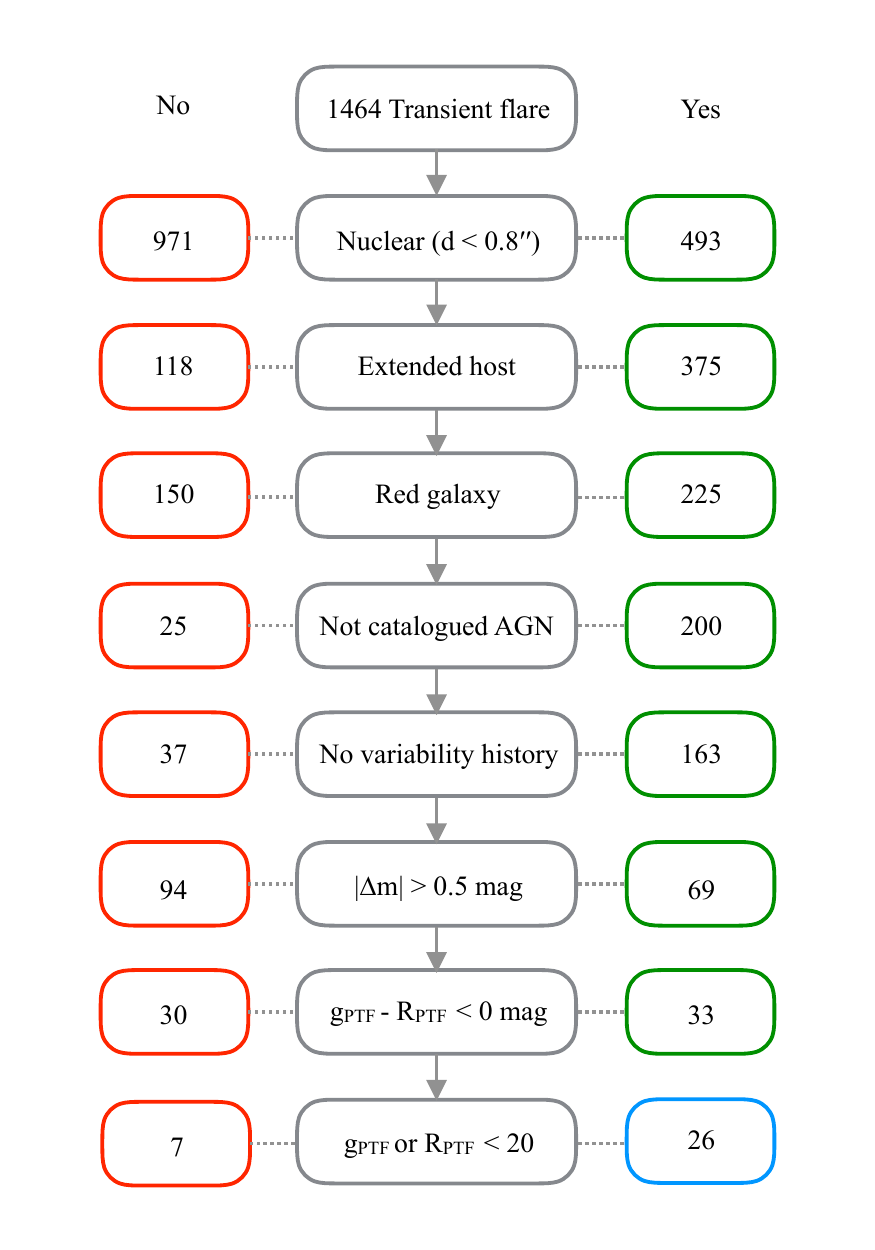}
\caption{Flow diagram of the filtering process for TDE candidates. The composition of each intermediate step is listed in \autoref{tab:selection}.}
\label{fig:flow}
\end{figure}

The filtering has 7 steps (\autoref{fig:flow}). We first select sources that are astrometrically aligned with the center of a morphologically extended host galaxy. We make a further cut based on host galaxy color, selecting only nuclear transients in red host galaxies. We then remove sources that are known AGN by cross-matching with external catalogs or checking for flaring history in the historical PTF and iPTF database since 2009. We exclude sources with weak variability by applying a magnitude cut. Lastly, we remove red transient events and sources that are too dim for spectroscopic follow up. The composition of each intermediate sample is listed in \autoref{tab:selection}. The first column contains the description of each filtering procedure, which is detailed in the following sections.

\begin{deluxetable*}{lccccccccccc}
\tablecaption{Sample composition after each filtering step\label{tab:selection}}
\tablecolumns{12}
\centering
\tablehead{\colhead{Selection} & \colhead{all} &\multicolumn{2}{c}{Host color} & \colhead{Known} & \colhead{Historical}
& \multicolumn{2}{c}{$|\Delta m_{var}|$} & \multicolumn{2}{c}{\gptf{}$-$\rptf{}} & \multicolumn{2}{c}{Peak mag} \\
\colhead{} &\colhead{} &\colhead{blue} &\colhead{red} &\colhead{AGN} &\colhead{Detection} &\colhead{$\geq$0.5} &
\colhead{$\leq$0.5} &\colhead{$\leq$0} &\colhead{$\geq$0} &\colhead{$\leq$20} &\colhead{$\geq$ 20}}
\startdata
Nuclear & 493		& 216 	  & 267 & 37 		& 94 		& 249 	& 235 		& 200 	& 203 		& 285 & 206 \\
Extended host & 375		& 150	  & 225 & 31 		& 62 		& 189 	& 186 		& 158 	& 157 		& 221 & 154 \\
Red host & 225		& \nodata & 225 & 12 		& 37 		& 81 	& 144 		& 78 	& 105 		& 136 & 89 \\
Not known AGN & 200		& \nodata & 200 & \nodata	& 37 		& 78 	& 122 		& 72 	& 95 		& 119 & 81 \\
No variability history & 163		& \nodata & 163 & \nodata 	& \nodata 	& 69 	& 94 		& 62 	& 73 		& 90 & 73 \\
$|\Delta m_{var}|$ $>$ 0.5 & 69		& \nodata & 69 	& \nodata 	& \nodata 	& 69 	& \nodata 	& 33 	& 30 		& 46 & 23 \\
Blue flare & 33		& \nodata & 33 	& \nodata 	& \nodata 	& 33 	& \nodata 	& 33 	& \nodata 	& 26 & 7
\enddata
\end{deluxetable*}

\subsection{Selection of Nuclear Transients (Step 1)}
\label{subsec:step1(nuclear)}
Tidal disruption of a star around a black hole occurs when the orbit of the star crosses the tidal disruption radius ($R_T$). For a non-spinning black hole, $R_T$ is defined as
\begin{equation}
R_T = R_\star \left(\frac{M_{BH}}{M_\star}\right)^{1/3} = 0.47 AU \left(\frac{R_\star}{R_\odot}\right) \left(\frac{M_{BH}}{10^6 M_\star}\right)^{1/3},
\end{equation}

where $M_\textrm{BH}$ is the black hole mass, $R_\star$ and $M_\star$ are the radius and mass of the disrupted star, respectively. Since a TDE happens near the black hole, we limit our search to transients that reside in the center of a host galaxy. This tidal disruption radius corresponds to an angular separation on the nano-arcsecond scale at $z$=0.1 that cannot be resolved by any instrument.

\begin{figure}
\centering
\includegraphics[width=3.5in, angle=0]{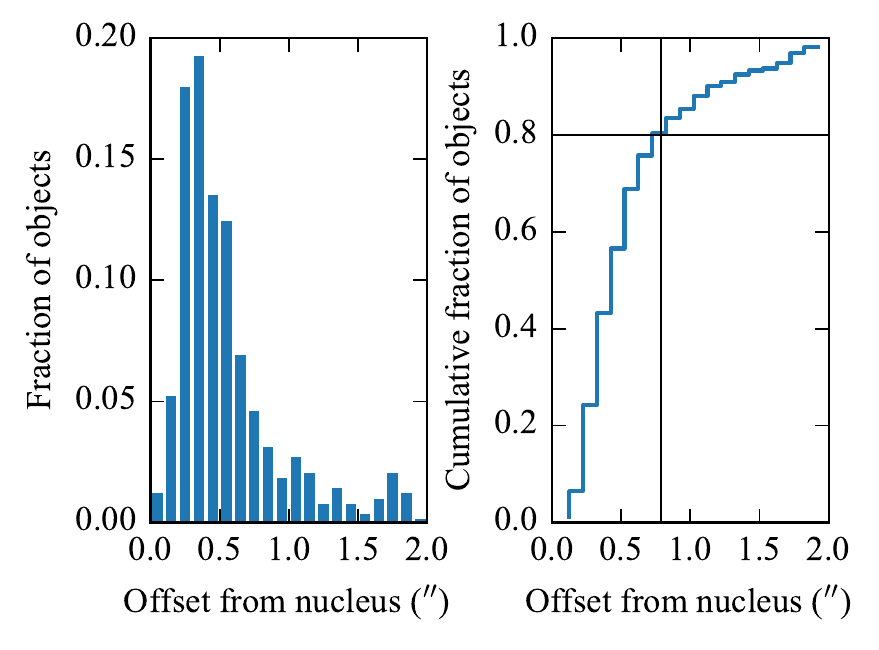}
\caption{Flare-host separation for transients associated with known AGNs in PTF database (2009--2012). The spatial cut at $d <$ 0.8\arcsec{} includes 80\% of the AGNs.}
\label{fig:agnoffset}
\end{figure}

We define a transient as nuclear if the separation of the transient and the centroid of the galaxy is less than 0.8\arcsec. This radial cut was empirically determined to include 80\% of the PTF transients in the database associated with known AGNs (\autoref{fig:agnoffset}). This offset distance is measured relative to the coordinates of the host centroid as output by \texttt{SExtractor} \citep{2016PASP..128k4502C}. While the mean position of the offset (relative right ascension $\Delta\alpha$ and declination $\Delta\delta$) over all detections should be zero, the offset distance between the transient and the host $x\approx$$\sqrt{\Delta\alpha^2 + \Delta\delta^2}$, should be non-zero (\autoref{fig:agnoffset}). The offset distance distribution can provide clues to the distribution of $\Delta\alpha$ and $\Delta\delta$. For example, if both $\Delta\alpha$ and $\Delta\delta$ are independent and normally distributed with the same variance and zero means, the offset distances can be characterized by a Rayleigh distribution that has the following form.

\begin{equation}
f(x;\sigma) = \frac{x}{\sigma^2}e^{-x^2/2\sigma^2},
\end{equation}

where $x$ is the offset distance and $\sigma$ is the standard deviation of the $\Delta\alpha$ and $\Delta\delta$ distributions. The mode (0.4\arcsec{} in \autoref{fig:agnoffset}) of a Rayleigh distribution is equal to 1$\sigma$.

Cross-matching to AGNs is intended to help us quantify the tolerance for scatter in offset distance for nuclear transients. We expect some of these matched transients to be caused by unidentified supernovae instead of nuclear activity. In fact, the tail of the offset distance distribution should be mostly contributed by these supernovae that went off near AGNs.

For each object, we calculate the median of the separation of all detections available at the time a transient was saved and apply the separation cut to remove 971 transients that are non-nuclear. On average, P48 detected 6 new nuclear transients and 12 non-nuclear transients per night for an average nightly coverage of 771 deg$^2$. This corresponds to a surface density of 7.8$\times$10$^{-3}$ per deg$^{2}$ for nuclear transients on the sky.

\subsection{Selection of Extended Hosts (Step 2)}
\label{subsec:step2(ext host)}
We remove transients with unresolved hosts that are classified as a star or a quasar. We searched for host galaxies within a 3\arcsec{} radius of a transient using SDSS DR12 and PS1 DR1 catalogs. The SDSS star-galaxy separator comes with the \texttt{frames} pipeline and the results are stored in \texttt{type} under the \texttt{PhotoObjAll} table. In cases where a source is outside the imaging coverage of SDSS, we use measurements from PS1 3$\pi$ Survey, which is available in DR1, to remove point-like sources. We define a point source as having \texttt{rMeanPSFMag}~$-$~\texttt{rMeanKronMag} $<$ 0. This definition can segregate stars and galaxies with a 99\% accuracy for galaxies brighter than 20 mag in the $r$ band (Tachibana et al. in prep).

\subsection{Selection of Red Galaxy Hosts (Step 3)}
\label{subsec:step3(host color)}
\begin{figure}
\centering
\includegraphics[width=3.5in, angle=0]{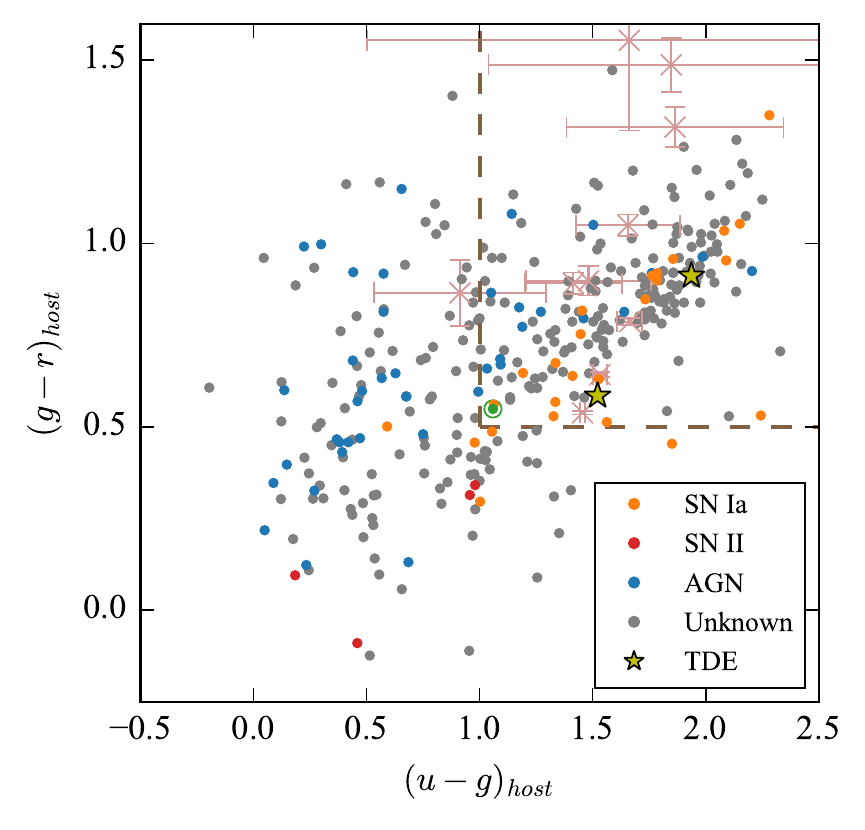}
\caption{Color-color diagram of the host galaxies of iPTF nuclear transients. An empirical host galaxy color cut for red galaxies is represented by the dashed lines, where sources located in the top right corner satisfy our selection. The pink crosses show the host galaxy colors of previously reported TDEs from four optical surveys: PS1, PTF, SDSS, and ASASSN (see selection of comparison in \autoref{sub:completeness}) that are inside SDSS footprint while the two stars mark the two TDEs (iPTF16axa and iPTF16fnl) discovered during this experiment. The only TDE located outside of this dashed box is PS1-10jh. However, a deeper imaging of PS1-10jh does show it satisfying this color selection; it is slightly outside the bounding box because of the high uncertainty in SDSS $u$ photometry. By employing this color cut, we can filter out the CCSNe (red), which are preferentially hosted by blue starforming galaxies as well as a large fraction of AGNs. The green circle marks PS16dtm, which was reported as a TDE in a NLS1 galaxy.}
\label{fig:hostcolor}
\end{figure}

The major sources of contamination in the nuclear transient sample are SNe that fall inside the nuclear circle of a host galaxy and variable AGNs. While previously discovered TDEs seem to be found toward early-type galaxies, it has been well-established that core-collapse supernovae (CCSNe) are preferentially hosted by "blue" starforming galaxies \citep[e.g.][]{2012ApJ...759..107K}. Therefore, we select only nuclear events hosted by red galaxies with empirically defined boundaries of $u-g>$ 1.0 mag and $g-r>$ 0.5 mag. This definition has worked well for previously discovered TDEs, which are shown as crosses in pink in \autoref{fig:hostcolor}. For transients outside of SDSS footprint, we use a looser host galaxy color cut that only requires $g-r>$ 0.5 mag using PS1 photometry since the $u$ filter is absent in PS1.

The $u-g$ and $g-r$ colors of the SDSS host galaxies of the nuclear events are shown in \autoref{fig:hostcolor}.
Four CCSNe have been identified with follow up spectroscopy and the colors of their host galaxies are marked in red. As expected, the host galaxies of all four CCSNe have blue colors and are excluded by our empirical host color cut. Transients that are associated with known AGN hosts or classified with follow up spectroscopy are labeled in blue. The host galaxy color cut removes 27 out of 41 (66\%) known or newly typed AGNs since a strong blue continuum is usually present in AGN spectra. In this step, we have removed iPTF16bco, which was found to be a changing-look quasar with a continuum flux increase of at least a factor of 10 over $\lesssim$ 1 yr \citep{2017ApJ...835..144G}.

\subsection{Removal of AGNs (Step 4)}
There is no physical process restraining TDEs from occurring in an AGN. However, AGNs are much more common (100x) than TDEs in the Universe and most of them exhibit some extent of intrinsic variability that is not of the interest of this study. To optimize the usage of follow up resources, we choose to eliminate transients that are likely due to AGN variability. We first remove known AGNs from the sample of TDE candidates. The candidate sample from step 3 is cross-matched with the 13th edition of quasars and active nuclei catalog \citep{2010A&A...518A..10V}. In addition, we also search SDSS for spectroscopic classification labels in table \texttt{SpecObj} and \texttt{galSpecInfo}. Nuclear events with host galaxies labelled as \texttt{QSO}, and \texttt{AGN} are most likely due to intrinsic variability and are removed to narrow down the sample size of candidate TDEs.

Secondly, we remove sources with variability history in PTF and iPTF from the sample since variability is an ubiquitous signature of AGN \citep[e.g.][]{VandenBerk2004,Wilhite2005}. We search for detection history in archival PTF and iPTF database from 2009 to 2016 at the coordinates of the nuclear transients. To avoid dubious subtractions, we require at least two detections ($>$5$\sigma$) on the same night to report the detection as 'real'. A source is removed from the sample if it was detected in at least one of the previous seasons and was not reported as any transient phenomenon other than AGN.

By limiting the search to inactive galaxies only, we have eliminated iPTF16ezh (or PS16dtm) from the sample, which is a unique event that has been reported as a TDE in a narrow line Seyfert 1 galaxy at $z$ = 0.08 by \cite{2017ApJ...843..106B}.

\subsection{Variability Amplitude Cut (Step 5)}
We exclude sources with insignificant variability amplitude by comparing the magnitude of the flare with respect to the PSF magnitude of the host. We define the the amplitude of variability ($\Delta m_{var}$) as
\begin{equation}
\Delta m_{var} = -2.5\textrm{log}_{10}(10^{-\frac{m_\textrm{PSF}}{2.5}}+10^{-\frac{m_\textrm{trans}}{2.5}})-m_\textrm{PSF},
\end{equation}
where $m_\textrm{PSF}$ is the PSF magnitude of the host galaxy and $m_\textrm{trans}$ is the host-subtracted transient magnitude. We remove nuclear transients with $|\Delta m_{var}| < $ 0.5 mag that can be attributed to small variability from AGNs. By imposing the 0.5 mag variability cut, we are only sensitive to sources with a $>$60\% flux increase in the nuclear region. This parameter can also be considered as a selection for high contrast against the core of the galaxy.

\subsection{Selection of Blue Flares (Step 6)}
Since the spectral energy distribution (SED) of optically bright TDEs can be approximated by a blackbody peaking in the UV, these events are intrinsically blue. Therefore, we use \gptf{}$-$\rptf{} colors from P48 and P60 to remove redder events that are most likely to be SNe. We query the database to calculate the mean \gptf{}$-$\rptf{} color over a timespan of one week from the date of discovery for each source. We correct for Galactic extinction using the Cardelli et al. (1989)
extinction curve with $R_C$ = 3.1 and $E(B-V)$ values based on the \cite{2011ApJ...737..103S} dust map. We do not take the K-correction into account when calculating the color of the flares. Since a TDE spectrum is dominated by a hot blackbody with the optical portion being on the Rayleigh-Jeans tail, the color is not significantly affected by the K-correction. Considering a case where the K-correction would have a greater impact, we assume a blackbody spectrum with $T_{BB}$=2$\times$10$^4$~K, which is on the cooler end for TDEs. At $z$=0.3, the $g-R$ color would recover from $-$0.26 to $-$0.36 after applying K-correction. The correction on $g-R$ color would be even smaller for TDEs with typical temperature (3$\times$10$^4$~K) or at a lower redshift.

\begin{figure}
\centering
\includegraphics[width=3.5in, angle=0]{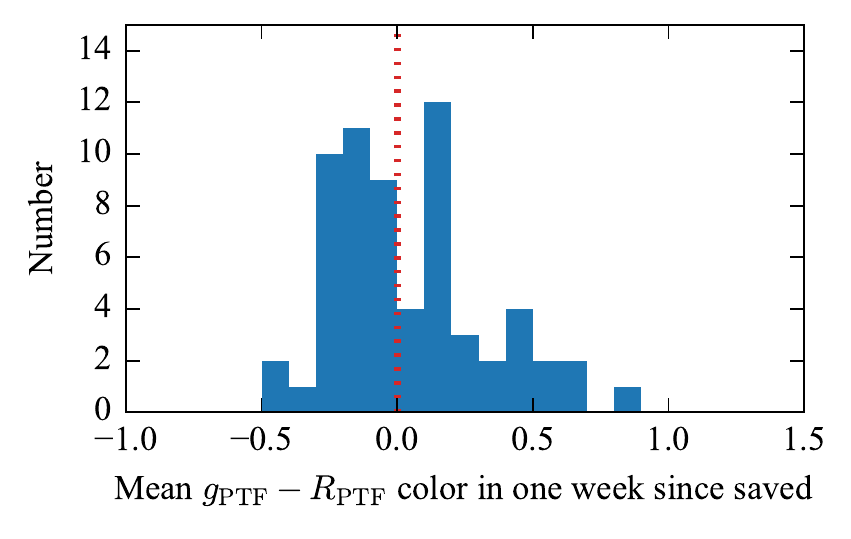}
\caption{The average \gptf{}$-$\rptf{} color of the transients in the first week of discovery. The colors shown here are Galaxy extinction corrected. With the color cut \gptf{}$-$\rptf{}$<$ 0 mag, we removed 30/63 ($\approx$48\%) of the sources.}
\label{fig:gr}
\end{figure}

An important caveat is, however, the host extinction, which is expected to be non-negligible in galaxy nuclei. This value is largely unknown and sometimes may result in the rejection of a transient with a color cut. A steep extinction curve like SMC would redden the \gptf{}$-$\rptf{} color by 0.12 mag. Considering these possible reddening mechanisms, we chose a conservative blue flare cut at \gptf{}$-$\rptf{}$=$0. We only use \gptf{} and \rptf{} pairs that were observed on the same night with the same telescope (P48 or P60) for the color calculation. With the condition \gptf{}$-$\rptf{}$<$0 mag (\autoref{fig:gr}), roughly half of the flares were rejected from the sample.

\subsection{Brightness Limit for Spectroscopic Follow-up (Step 7)}
Finally, we remove sources that were too dim for spectroscopic follow up with 4-meter class telescopes (e.g. DCT) if \gptf{} or \rptf{} did not reach $<$20 mag in 3 days after being saved. The final sample consists of 26 sources.

\section{Photometric and Spectroscopic Follow up}
\label{sec:observations}
\subsection{P48 and P60 photometry}
The Palomar 48-inch telescope (P48) is the workhorse of iPTF. During the experiment, the targeted fields were exposed for 60s for each image to reach a depth of $\sim$20.5 mag in the \gptf{} and \rptf{} bands. The raw images were detrended and astrometrically and photometrically calibrated at IPAC \citep{2014PASP..126..674L}.

Follow-up photometric observations with the robotic Palomar 60-inch telescope (P60) was sometimes triggered in $gri$ bands to get color and better sampling of the light curve. The P60 raw images were processed by the Fremling Automated Pipeline to be subtracted against the reference images from SDSS \citep{2016A&A...593A..68F}. We did not obtain any host-subtracted photometry for sources outside of the SDSS footprint with P60.

\subsection{\Swift{} UVOT and XRT photometry}
We triggered Neil Gehrels \Swift{} \citep{gcg+04} UVOT \citep{2005SSRv..120...95R} observations in the bluest $uvw2$ (1928 \AA) filter from our Cycle 12 Key Project (PI: Gezari) for the classification of iPTF nuclear transients. The exposure time for each \Swift{} trigger is 1~ks in order to achieve a limiting magnitude of $uvw2\sim$~23 mag. We used the task \texttt{uvotsource} in the HEASoft package to extract the photometry within a 5\arcsec{} radius aperture in the AB magnitude system.

Simultaneous \Swift{} XRT \citep{bhn+05} observations were processed by the Swift Data Science Centre pipeline. The pipeline extracts counts from the source in the energy band 0.3 -- 10 keV by accounting for dead columns and vignetting.

\subsection{SED Machine}
The Spectral Energy Distribution Machine (SEDm) is a integral-field-unit (IFU) spectrograph mounted on the P60 telescope \citep{2017arXiv171002917B}. The wide FOV (28\arcsec{}), low resolution (R$\sim$100) IFU is a powerful tool for classifying transients down to $r \sim 19$ mag. The data obtained by SEDm IFU were automatically processed by the data reduction pipeline, which includes basic image reduction, defining the IFU spatial and wavelength geometry, and spectral extraction. The extracted spectra were flux calibrated with the observations of spectrophotometric standard stars.
\subsection{P200 DBSP}
We observed with the Double Beam Spectrograph (DBSP) on the 200-inch Hale telescope at Palomar Observatory (P200). A dichroic is used to split the light into a blue and a red component. The observing setup includes a 1\arcsec{} wide slit and 600g~mm$^{-1}$ grating, giving a dispersion of 1.5 \AA~pixel$^{-1}$. The images were reduced with the \texttt{pyraf-dbsp} script.
\subsection{DCT}
Spectroscopic observations were made with the DeVeny spectrograph mounted on the 4.3-meter Discovery Channel Telescope at Lowell observatory. We used a slit width of 1.5" since the seeing did not get better than 1\arcsec{}. A 300g~mm$^{-1}$ grating was used to achieve a dispersion of 2.2 \AA~pixel$^{-1}$. We used standard IRAF routines to reduce the data. The procedures include bias subtraction, flatfielding, aperture extraction, wavelength calibration, and flux calibration.
\subsection{Keck LRIS}
Spectroscopic observations were made with the Low Resolution Imaging Spectrometer \citep[LRIS;][]{1994SPIE.2198..178O} on Keck observatory. The observing configuration included a 1\arcsec{} slit and a 400/3400 grism that gives a FWHM resolution of $\sim$7\AA. The data were reduced with the LRIS automated reduction pipeline\footnote{\url{http://www.astro.caltech.edu/~dperley/programs/lpipe.html}} and flux-calibrated with the standard star BD+28d4211.

\section{Classification}
\label{sec:classification}
The photometric properties of the final sample of 26 candidates are listed in \autoref{tab:sample}. Classification spectra were taken for 9 of the candidates. We requested \Swift{} UVOT photometry in $uvw2$ band for 7 of the high-confidence transients in this final sample that did not have light curves or color evolution resembling a SN Ia or AGN, and did not have a classification spectrum. We describe the classification based on both spectroscopy and photometry in the following sections. Even if a source was not followed-up during the flare, we still obtain spectroscopy of their host galaxies at a later time to measure their redshifts. The host spectra are shown in the Appendix with observing details described in \autoref{tab:host_obs}. The spectroscopic data will be be made available via WISeREP \citep{2012PASP..124..668Y}.

\subsection{Supernovae}
\subsubsection{Type Ia Supernovae}
The similarity in SN Ia light curves has direct application to determining cosmological distances. The SN Ia light curves become even more uniform after accounting for the empirical correlation between the maximum luminosity and the width of the light curve \citep{1993ApJ...413L.105P}, making them the standard candles in the Universe. Unlike CCSNe, SNe Ia do not have a preference for star-forming regions.

In this sample, 5 transients are spectroscopically confirmed SNe Ia classified with the Supernova Identification code \citep[SNID;][]{2007ApJ...666.1024B}. For the transients without any follow up spectrum, we inspect the light curves by eye and found 9 transients having burst-like light curves and/or \gptf{}$-$\rptf{} color reddening with time that are characteristic of a SN Ia. We label them as ``SN Ia phot'' in \autoref{tab:sample}. A typical SN Ia light curve is characterized by a smooth rise ($\lesssim$20 days) and decay ($\sim$0.1 mag per day). Contrary to the lack of color evolution in TDEs, the color of a SN Ia reddens as the ejecta expand with time.

We obtained host spectra for all of these photometric SNe Ia (\autoref{fig:SNI}) in 2017 and measured their redshifts. The peak absolute magnitudes of these photometric SNe Ia are all typical of a SN Ia with m$_{peak}$ around $-$19 mag. \Swift{} observations in $uvw2$ for iPTF16fmd and iPTF16gyl were both $>$22 mag, which is consistent with the SN Ia classification.

\begin{figure}
\centering
\includegraphics[width=3.5in, angle=0]{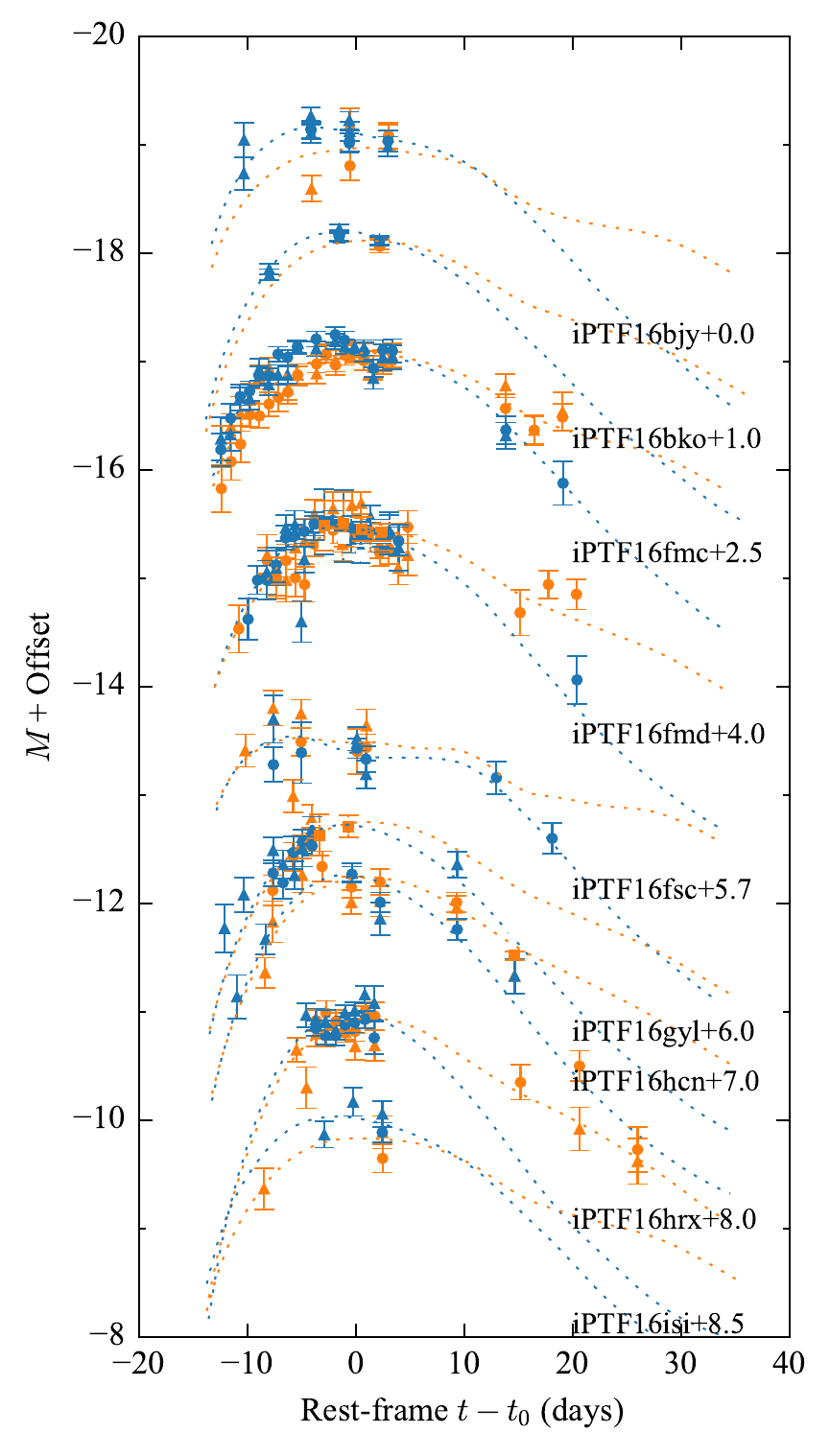}
\caption{Light curves of the photometric SN Ia in the sample. The blue and orange dotted curves show the best-fit result for \gptf{} and \rptf{} band data with the SALT2 model. The circles show the photometry extracted by the IPAC pipeline while the triangles show the photometry extracted by the NERSC pipeline. }
\label{fig:SN_fit}
\end{figure}

To verify our classification, we perform light curve fitting on all 9 photometric SNe Ia as well as 5 spectroscopically confirmed SNe Ia using the implementation of the SALT2 model in Python package \texttt{sncosmo}\footnote{\url{https://github.com/sncosmo/sncosmo}}. By fixing the redshift determined from host galaxy spectra, we simultaneously fit light curve data points from both IPAC and NERSC pipelines with the SN Ia time-series model convolved with filter response in \gptf{} and \rptf{}. The key parameters in this model are the date of $B$--band maximum ($t_0$), normalization $x_0$, light curve stretch $x_1$, and the color $c$. The best-fit light curves of the 8 photometric SN Ia are shown in \autoref{fig:SN_fit} while the best-fit parameters of all 14 sources are listed in \autoref{tab:SALT2_params}. Most of the sources (12/14) have a goodness of fit $\chi^2_\nu <$ 3.0. The 2 sources with a slightly larger $\chi^2_\nu$ include 1 spectroscopic (iPTF16bke) and 1 photometric (iPTF16hcn) SN Ia.

\subsubsection{Probable Core-Collapse Supernovae}

\begin{figure}
\centering
\includegraphics[width=3.0in, angle=0]{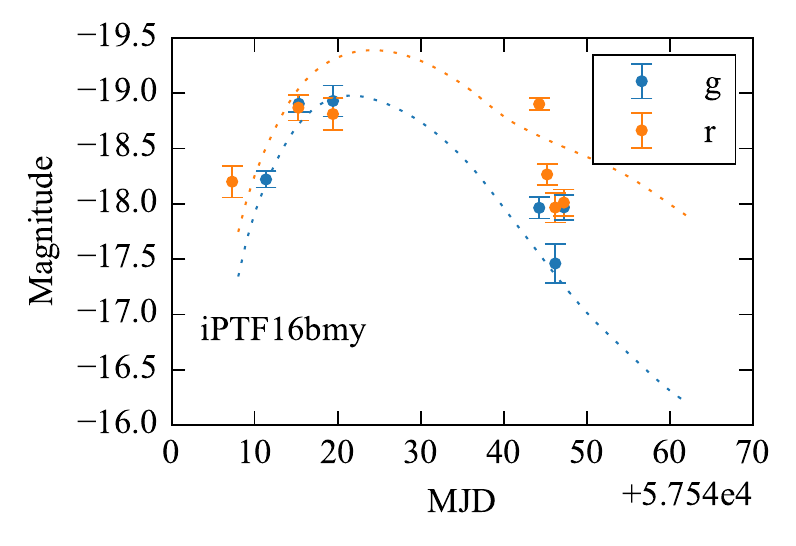}
\caption{The light curve of iPTF16bmy. The dotted line shows the best-fit SN Ia light curve with $\chi^2_\nu$ = 58.8. }
\label{fig:unclassified}
\end{figure}

While confirming the SN Ia classification through light curve fitting, we noticed an outlier, iPTF16bmy, that has a decline too steep to be consistent with a SN Ia light curve. We show the light curve of iPTF16bmy and the best-fit SN Ia light curve ($\chi^2_\nu$ = 58.8) in dotted lines in \autoref{fig:unclassified}.

We obtained a host spectrum of iPTF16bmy with P200 on Jul 29 2017. The host spectrum of iPTF16bmy (\autoref{fig:bmyhost}) shows absorption lines dominated by an old stellar population in an elliptical galaxy at $z$ = 0.129. The redshift corresponds to a peak magnitude of $-$18.9 mag in \gptf{}. The fast rise to peak ($<$10 days) would also be unusual for an AGN. The faster rise than decline timescales and the blue color (\gptf{}$-$\rptf{} = $-$ 0.14) near peak resemble the sample of rapidly evolving transients found by \cite{2014ApJ...794...23D}, which is likely to have a CCSN origin. All the host galaxies of the rapidly evolving transients have nebular emission lines that are indictive of ongoing star formation. By doing a careful fitting with \texttt{ppxf} (\autoref{fig:bmyhost}), we do see some weak level of H$\alpha$, [NII], and [OIII] emission. With the information available to us, we classify iPTF16bmy as a ``probable'' CCSN in \autoref{tab:sample} even though we cannot rule out iPTF16bmy as a subtype of SNIa.

\subsection{Variable AGN}

We obtained spectra of iPTF16ayd and iPTF16fqa during their flaring states and classified them as AGNs. iPTF16ayd was typed as an AGN due to the presence of the broad Balmer emission lines (FWHM(${H\beta}$) = 3080 km s$^{-1}$) that are typical of a Type 1 Seyfert. Although the light curve of iPTF16fqa surged by $\approx$2 mag in 60 days, its flaring spectrum does not show broad P-Cygni spectral features like an SN Ia. The \Swift{} observation of iPTF16fqa showed no X-ray emission though the $uvw2$ magnitude had increased by 1 mag compared to the pre-flare GALEX $NUV$ photometry. While the lack of P-Cygni line and increase in UV flux are both in favor of a TDE origin, we did not notice any \HeII{} emission that are often seen in optical TDEs in its flaring spectrum. In addition, we did not detect any change in the optical continuum or the emission line in this source when we made another spectroscopic observation with P200 in 2017. This is at odds with previously detected TDEs, where spectral evolution becomes noticable on weekly to monthly timescales \citep[e.g.][]{Holoien2014,2017ApJ...843..106B}. The narrow line ratios (\autoref{fig:BPT}) also place this object within the boundary for Seyfert galaxies. The overall spectroscopic properties of iPTF16fqa are more consistent with the scenario of a variable AGN.

\begin{figure*}
\centering
\includegraphics[width=7.0in, angle=0]{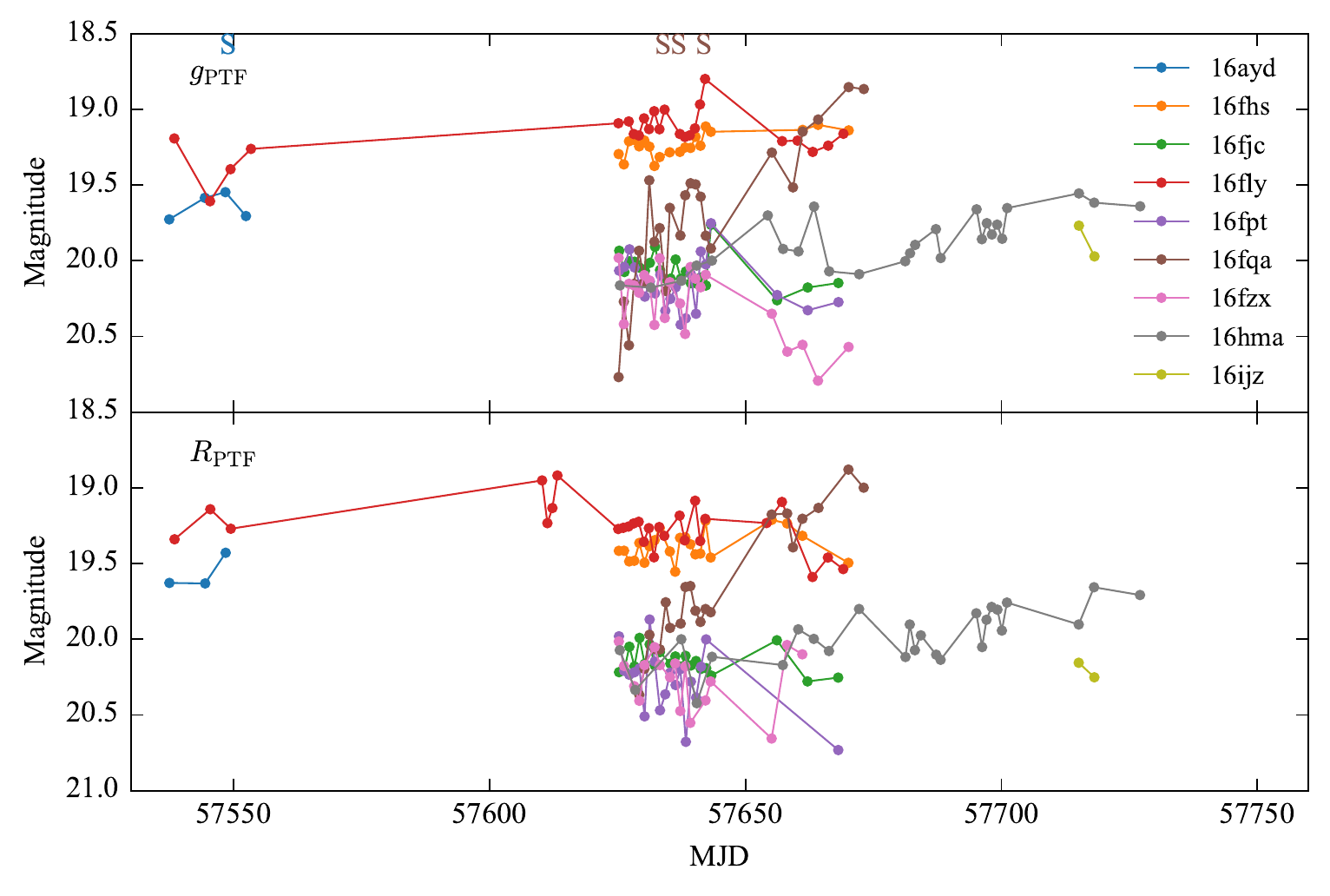}
\caption{Light curves of the spectroscopically confirmed AGNs and candidate AGNs in our sample. AGN candidates are selected based on their light curves, which lacks obvious rise or fall over the monitoring period. Archival PTF data shows no detection at the positions of the AGN candidates.}
\label{fig:AGN_LC}
\end{figure*}

\begin{figure}
\centering
\includegraphics[width=3.5in, angle=0]{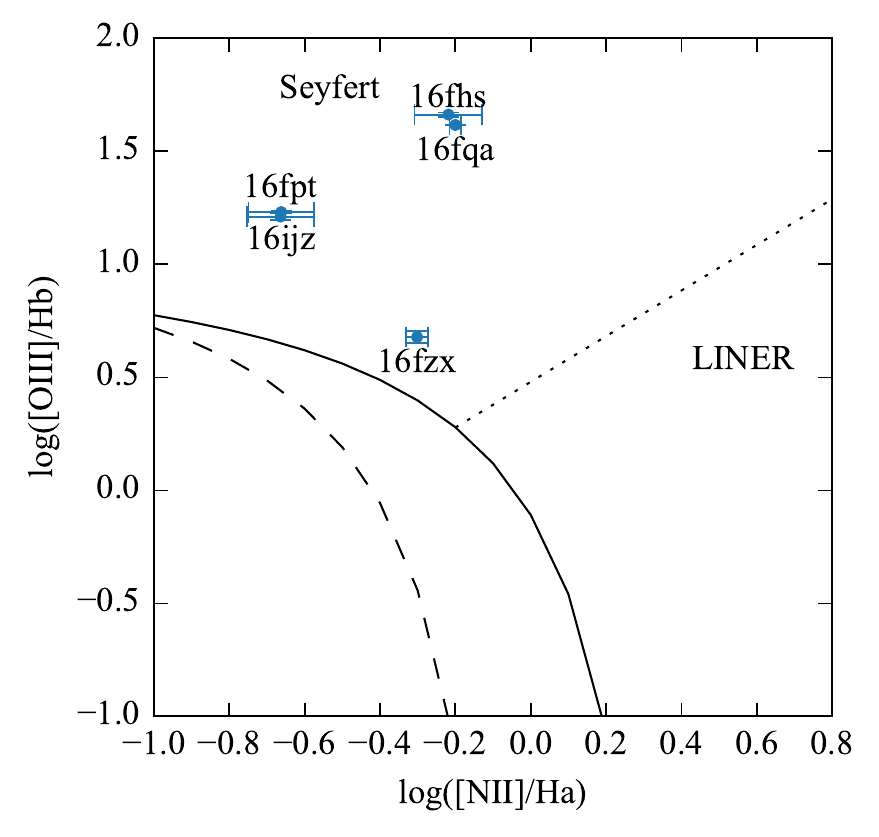}
\caption{BPT diagram with the solid \citep{2001ApJ...556..121K} and dashed \citep{2003MNRAS.346.1055K} line separating AGN from starforming galaxies. The dotted line represents the Seyfert--LINER demarcation \citep{2010MNRAS.403.1036C}. The photometric AGNs that do not have broad Balmer lines are color-coded in blue.}
\label{fig:BPT}
\end{figure}

We classify 7 other transients in the sample as photometric AGN because their light curves do not have an obvious rising or declining trend. The light curves of all 9 spectroscopic and photometric AGNs in our sample are shown in \autoref{fig:AGN_LC}.

We verify the photometric AGN classification with the host spectra obtained in 2017. Among these 7 photometric AGNs, iPTF16fly, iPTF16fjc, and iPTF16hma have broad Balmer emission lines (\autoref{fig:broadlineAGN}). We measure and report the broad H$\beta$ line widths in \autoref{tab:host_obs}. Among the sources in our sample, iPTF16fly has the highest redshift ($z$=0.45) that the H$\alpha$ is shifted out of the bandpass.
We measure the narrow line ratios and plot them on the [OIII]/H$\beta$ versus [NII]/H$\alpha$ diagnostic diagram (\autoref{fig:BPT}) for AGNs without broad emission lines. Their spectra are shown in \autoref{fig:type2AGN}. One of these transients, iPTF16fpt, has a pre-flare SDSS spectrum. Because the host spectrum of iPTF16fpt is dominated by the underlying galaxy, we re-measure the line ratios by fitting the SDSS spectrum with \texttt{ppxf} \citep{2017MNRAS.466..798C}, which allows simultaneous fitting of the emission lines and the host galaxy template.  We classify iPTF16fhs, iPTF16fpt, iPTF16ijz, and iPTF16fzx as AGNs based on their narrow line ratios as shown in \autoref{fig:BPT}.

All 9 transients that are classified as AGNs have a variability amplitude $|\Delta m_{var}| > $ 0.7 mag at peak. Five of them (iPTF16ayd, iPTF16fjc, iPTF16fly, iPTF16fpt, and iPTF16hma) have $|\Delta m_{var}| > $ 1.0 mag that satisfies the photometric selection critereon for changing-look quasar (CLQ) candidates in \citep{2016MNRAS.457..389M}. The flaring spectra for all four objects except iPTF16fpt, which only has a pre-flare SDSS spectrum, exhibit broad Balmer lines. Although we cannot verify whether these four objects have gone through a transition in spectral shape due to the absence of pre-flare spectra, it may be worthwhile to keep an eye on these candidates since some CLQs have shown a reversion to the original spectral class after tens of years \citep[e.g.][]{2016A&A...593L...8M}.

CLQs are a recently emerged class of objects that exhibit extreme variability in both continuum and emission lines on the timescale of a few years \citep{2014ApJ...788...48S,2015ApJ...800..144L,2017ApJ...835..144G}. The name refers to the fact that their optical spectra change from a Type 1 to a Type 2 Seyfert or vice versa. Compared to mundane AGNs that only vary by 20\% on timescales of months to years \citep[e.g.][]{1994MNRAS.268..305H}, CLQs have sparked even more questions on the origin of AGN variability. Currently, it is generally agreed that variable obscuration in the line of sight, lensing of a background quasar, or a tidal disruption event cannot explain every aspect of a CLQ.

This leaves changes in the accretion rate as the favored mechanism for what is causing the CLQs. However, the timescale for viscous perturbations to propagate in a classical thin accretion disk in the UV/optical emitting region ($\approx$ 100$r_g$) is on the order of 10$^3$ years \citep{2016ApJ...833..226H}, which is much longer than what was observed for CLQs. But this may be resolved if the UV/optical emission of CLQs are irradiated spectrum with variability originating from a smaller radius, where the viscous timescale is effectively reduced.

It has been shown that the difference spectra of quasars could be well fitted by a thin accretion disk spectrum with changing accretion rates \citep[e.g.][]{Pereyra2006,2016ApJ...833..226H}, which is consistent with a disk spectrum that follows $f_\nu \propto \nu^{1/3}$ in the optical emitting region. Even in the extreme case of CLQs, it has been reported in some cases that the difference spectra also have power law indices in agreement with a geometrically thin optically thick accretion disk \citep{Gezari2017,2016MNRAS.457..389M}.

The power-law thin disk spectrum corresponds to a \gptf{}$-$\rptf{}$\sim$ $-$0.1 mag. The \gptf{}$-$\rptf{} colors measured for iPTF16fhs, iPTF16fjc, iPTF16fpt, and iPTF16fzx are consistent with this value, while iPTF16fly, iPTF16fqa, iPTF16hma, and iPTF16ijz are slightly bluer with \gptf{}$-$\rptf{}$\sim$ $-$0.2 mag. It is interesting that 3 out of 4 Type 1 Seyferts with broad emission lines, including iPTF16ayd (\gptf{}$-$\rptf{}$\approx$ $-$0.01 mag), seem to deviate from the thin disk prediction. As discussed in \cite{2016ApJ...833..226H}, the deviation could be caused by the presence of broad emission lines in the bandpass that are also variable. While changes in accretion rate can describe the change in the AGN continuum quite well, variability in the broad emission lines may be driven by different physical mechanisms.

\subsection{TDEs}
During the rolling \gptf{}+\rptf{} experiment, iPTF discovered two TDEs, iPTF16axa at $z=$0.108 \citep{2017ApJ...842...29H} and iPTF16fnl at $z=$0.016 \citep{2017ApJ...844...46B}, on UT 2016 May 29 and 2016 August 29, respectively. \Swift{} was triggered immediately after the discovery for iPTF16axa. The brightening in the $uvw2$ filter along with the strong HeII and H$\alpha$ line in the Keck DEIMOS classification spectrum are both in agreement with the TDE interpretation.
iPTF16axa has a persistent blackbody temperature of $\sim$3$\times$10$^4$~K throughout the 3-month monitoring period. The light curve of iPTF16axa followed the classic $t^{-5/3}$ power-law decline. No X-ray or radio emission were detected for this source. In \cite{2017ApJ...842...29H}, we measured the stellar velocity dispersion with high resolution spectroscopy and derived a black hole mass of $M_{BH}$ = 5.0$^{+7.0}_{-2.9}\times$10$^6$ M$_\odot$ using the M--$\sigma$ relation in \cite{2013ApJ...764..184M}. The black hole mass and the overall light curve properties of this source resemble the archetypal TDE PS1-10jh \citep{Gezari2012}.

On the other hand, iPTF16fnl was classified on the same night of discovery by the low resolution SEDm \citep{2017arXiv171002917B}. iPTF16fnl is the nearest (66~Mpc) and also the faintest TDE ever found at optical wavelengths. The light curve of iPTF16fnl declined much more rapidly than the other optically detected TDEs. Whether this decline follows an exponential form or a power-law with a steeper exponent is still debated. iPTF16fnl also has a constant blackbody temperature of $\sim$2$\times$10$^4$~K and its host galaxy happens to be an E+A galaxy that features strong H$\delta$ absorption.

Both TDEs exhibit strong HeII and H$\alpha$ line in their spectra. The helium-to-hydrogen ratios in these two objects are both higher than what would be expected in a nebular environment with solar abundance of He/H. This suggests that optical depth effects may be important in the line emitting gas in TDEs. For example, high density gas could lead to the suppression of Balmer lines as these transitions become optically thick \citep{2017ApJ...842...29H,Roth2016}.

\subsection{The Swift Sample}
\label{subsec:swiftsample}
\begin{figure}
\centering
\includegraphics[width=3.5in, angle=0]{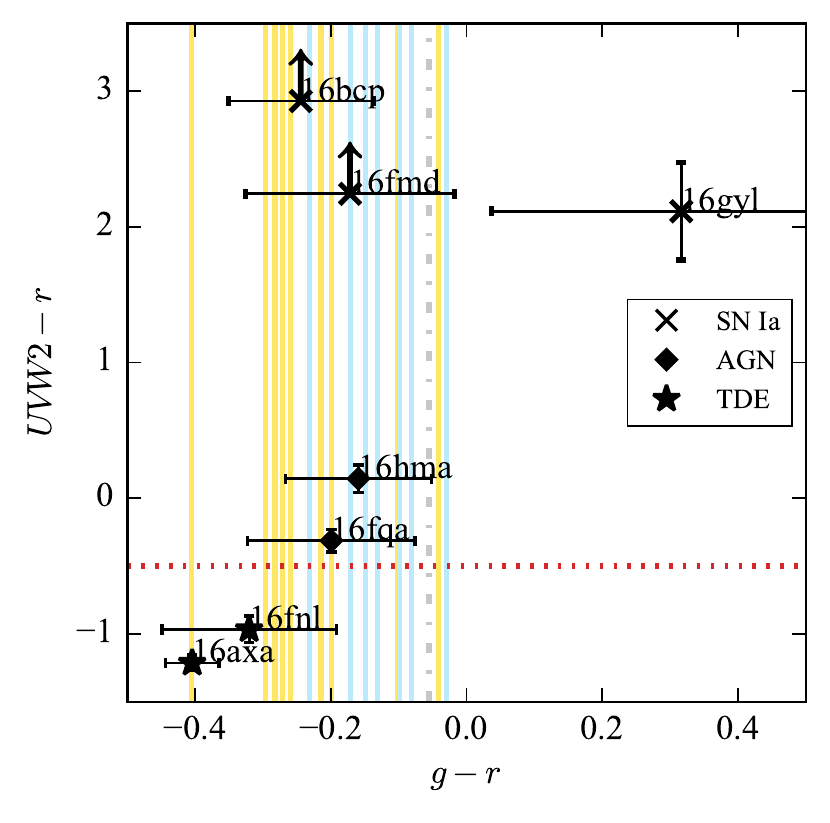}
\caption{Color-color diagram of the transient events in the sample. The symbols mark sources that were observed by \Swift{} and the \gptf{} and \rptf{} magnitudes are chosen to be the nearest to the time of the \Swift{} observation. The vertical lines show the mean \gptf{}$-$\rptf{} color of the sources in the final sample that were not observed by \Swift{}. The yellow lines represent sources that are classified as SNe Ia while the cyan lines represent sources that are classified as variable AGNs. iPTF16gyl had a \gptf{}$-$\rptf{} color of $-$0.3 when we triggered \Swift{} but has reddened in \gptf{}$-$\rptf{} by the time \Swift{} observed it.}
\label{fig:transcolor}
\end{figure}

We used \Swift{} ToO observations in the UV and X-rays under our Cycle 12 program (PI Gezari) to identify TDEs from our high-confidence candidates that do not have light curves resembling a SN Ia or an AGN at the time of discovery. We obtained \Swift{} observations for a total of 7 candidates in our sample, 6 of which were scheduled with the time allocated to our Cycle 12 program and 1 (iPTF16fnl) was scheduled through the regular ToO request. We triggered \Swift{} on these 6 candidates because we were unable to assign them to any spectroscopic observing runs in the timeframe of 3 days after saving. A separate \Swift{} ToO request for a two-months long monitoring campaign was submitted for iPTF16fnl after the rapid classification by SEDm. Details of these \Swift{} observations are presented in \autoref{tab:swiftsample}. We show the Galactic extinction corrected \gptf{}$-$\rptf{} and $uvw2-$\rptf{} color of all the nuclear events in our sample in \autoref{fig:transcolor}. The \gptf{} and \rptf{} photometry of the flares are chosen to be the nearest to the epoch of the \Swift{} observation. We note that although the $g-r$ color of iPTF16gyl had reddened when \Swift{} observed it, it had a \gptf{}$-$\rptf{} color of $-$0.3 at the time that we triggered \Swift{}. We also plot the mean \gptf{}$-$\rptf{} color in the first week of discovery for the AGNs (blue), SNe Ia (yellow), and unclassified sources (grey) in our sample. The two TDEs are well separated from the rest of the non-TDEs in $uvw2-$\rptf{}$<$$-$0.5 mag (red dotted line).

While we are able to get rid of the AGNs by adopting a more stringent optical color cut on the flare (\gptf{}$-$\rptf{}$<-$0.2 mag), SNe Ia can appear as blue as the TDEs up to one week post peak \citep[e.g.][]{2017ApJ...848...59M}. We therefore conclude that the UV is an important discriminator between TDEs and SNe Ia at early times.

\section{Discussion}
\label{sec:discussion}
\subsection{Selection on SDSS vs non-SDSS fields}
Among the unique fields, 77\%(509 out of 660) of the fields also lie within the survey footprint of SDSS DR12. The absence of SDSS data would have direct impact on the selection based on host galaxy properties (\autoref{subsec:step2(ext host)}, \autoref{subsec:step3(host color)}) since our selection requires $u-g >$~1.0 and $g-r >$~0.5 and the PS1 survey was conducted without the $u$ band. In our final sample, 19 out 26 sources were observed by SDSS and have archival SDSS photometry. If we apply the host morphology cut and host galaxy color cut with only $g-r >$ 0.5 for PS1 data, our pipeline would end up selecting 35 candidates. All nine additional candidates were rejected in our original selection for being in SDSS footprint but not satisfying $u-g >$~1.0 mag. From this, we estimate the contamination rate to increase by $\sim$35\% when only using PS1 data.

\subsection{Completeness}
\label{sub:completeness}

In order to determine the completeness of our search, we apply our selection criteria on 10 well-studied TDEs that were discovered by ground-based optical surveys. These 10 TDEs are TDE1 and TDE2 from SDSS \citep{vanVelzen2011}, PS1-11af \citep{Chornock2014}, PS1-10jh \citep{Gezari2012}, ASASSN-14ae, ASASSN-14li, ASASSN-15oi \citep{Holoien2014,Holoien2016,Holoien2016b}, PTF09ge, PTF09axc, and PTF09djl \citep{Arcavi2014}. We do not consider TDEs with DEC $<$ $-$20 deg (OGLE16aaa \citep{Wyrzykowski2016}, ASASSN-15oi \citep{Holoien2016b}, and ASASSN-15lh \citep{2016Sci...351..257D,2016NatAs...1E...2L}) here since we require host galaxies information from SDSS or PS1 in our selection.

All ten TDEs pass our selection criteria (see \autoref{tab: completeness_test}), with a few caveats described below. All ten TDEs have extended host galaxies and $g-r$ $<$ 0 at discovery. The only host galaxy the does not pass our host color cut is that of PS1-10jh, which has ($u-g$)$_{host}$ = 0.91$\pm$0.38 mag (\autoref{fig:hostcolor}) according to SDSS.
The large error bar in host color is mainly contributed by the faint $u$ band photometry with $u_{host}=$ 22.85$\pm$0.37 mag. In fact, the host color of PS-10jh does satisfy our criteria using a deeper stacked CHFT $u$ band and PS1 $g$ band image \citep{Kumar2015}, which indicates an (u-g)host = $2.00 \pm 0.015$ mag.

As mentioned earlier, PS16dtm is rejected by our selection since its pre-outburst archival SDSS spectrum indicates that the host is consistent with a Seyfert galaxy. All ten TDEs have $|\Delta m_{var}| > $ 0.5 mag at peak except TDE1, which has a $|\Delta m_{var}|$ of 0.3 mag. However, we must note that TDE1 was discovered after peak, and so it likely had a larger amplitude of variability. In sum, our selection criteria are inclusive of all the properties of all optically discovered TDEs hosted by quiescent galaxies.

\begin{deluxetable*}{lcccccc}
\tablecolumns{7}
\tablecaption{Applying selection criteria on previously reported TDEs. \label{tab: completeness_test}}
\tablehead{
\colhead{Name} & \colhead{ext} & \colhead{red host} & \colhead{Not AGN} & \colhead{No variability}
 & \colhead{$\Delta{m_{var}}$} & \colhead{$g-r$}}
\centering
\startdata

SDSS-TDE1 & Y & Y & Y & Y & -0.27$^{1}$ & -0.32 \\
SDSS-TDE2 & Y & Y & Y & Y & -0.73 & -0.31 \\
ASASSN-14ae & Y & Y & Y & Y & -0.54 & -0.27 \\
ASASSN-14li & Y & Y & Y & Y & -1.20 & -0.41$^{3}$ \\
PS16dtm & Y & Y & N & Y & -2.75 & -0.16 \\
PS1-11af & Y & Y & Y & Y & -0.98 & -0.30 \\
PS1-10jh & Y & Y$^{2}$ & Y & Y & -1.21 & -0.31 \\
PTF09ge & Y & Y & Y & Y & -1.25 & -0.25 \\
PTF09axc & Y & Y & Y & Y & -0.52 & -0.04$^{3}$ \\
PTF09djl & Y & Y & Y & Y & -0.97 & -0.24$^{3}$

\enddata
\tablenotetext{1}{Peak is not resolved in the light curve of TDE1.}
\tablenotetext{2}{Deeper $u$ band imaging confirms the host galaxy color of PS1-10jh satisfy $u-g >$~1 mag. }
\tablenotetext{3}{Since PTF09axc and PTF09djl only have single band light curve, we derive their observed $g-r$ by convolving a redshifted blackbody spectrum with the minimal best-fit temperatures reported in \cite{Arcavi2014}, with the \gptf{} and \rptf{} response functions, and adding in the Galactic extinction. The same is done for ASASSN-14li since the published $g$ and $r$ photometry include contribution from the host galaxy.}
\end{deluxetable*}

\subsection{Contamination}
\begin{figure*}
\centering
\includegraphics[width=7.0in, angle=0]{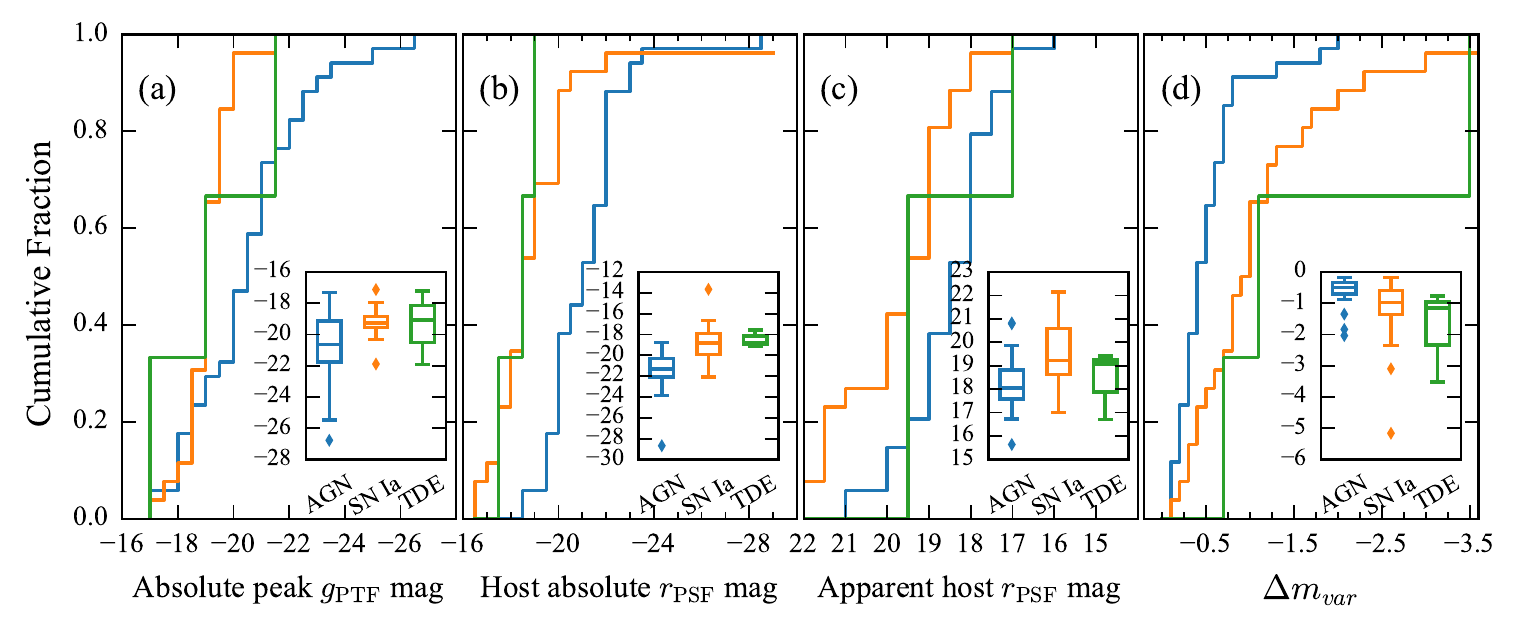}
\caption{The cumulative distribution of the (a) absolute peak magnitude in \gptf{} band, (b) absolute host PSF magnitude in $r$, (c) apparent host PSF magnitude in $r$, and (d) $\Delta m_{var}$ for spectroscopically typed nuclear transients with extended hosts. The insets show the distribution of parameters in the form of box plot that marks the interquartile range (IQR) for each object class. Although AGNs appear to be brighter than SNe Ia and TDEs in (a) and (b), it is uncertain if photometric redshift would be accurate enough for making these cuts. A magnitude cut on the host galaxy at apparent $r_\textrm{PSF} < $21.0 mag may help to remove some of the SNe Ia. Our variability amplitude cut can effectively remove AGNs as shown in (d).  }
\label{fig:cumulative}
\end{figure*}

Our final sample consists of 14 SNe Ia, 9 AGNs, 1 probable core-collapse SN, and 2 TDEs. To investigate if TDEs can be separated from AGN and SN in other parameter space, we use a control sample 26 SNe Ia, 34 AGNs, and 3 TDEs (including PS16dtm) to compare their properties. This control sample consists of spectroscopically classified sources derived from the nuclear transient sample ($d<$ 0.8\arcsec{}) in extended host galaxies in this study. We compare the cumulative distribution of different parameter space in \autoref{fig:cumulative} including the absolute peak magnitude in \gptf{} band, absolute host PSF magnitude in $r$, apparent host PSF magnitude in $r$, and $\Delta m_{var}$ in \autoref{fig:cumulative}.

As we pointed out in \autoref{subsec:swiftsample}, about 67\% (6 out of 9) of AGNs, 43\% (6 out of 14) of SNe Ia, and a probable CCSN can be effectively removed in our final sample by employing a more stringent flare color cut of \gptf{}$-$\rptf{}$<-$0.2 mag. We notice from panel b and c in \autoref{fig:cumulative} that AGNs have brighter absolute flare magnitudes and absolute host magnitudes compared to TDEs and SNe Ia, which would be two potential ways to remove AGN. Unfortunately, these two parameters require spectroscopy to determine the redshift and it is uncertain if the photometric redshift would be accurate enough for making these cuts. However, we find photometric criteria that can be applied to further reduce the contamination by AGNs and SNe Ia.  We verify that $|\Delta m_{var}| < $ 0.5 mag (panel d in \autoref{fig:cumulative}) is an effective filter for AGNs.  We also find that a magnitude cut on the host galaxy of $r_\textrm{PSF}$ $<$ 21.0 mag removes 25\% of the SNe Ia. This is in good agreement with the results of \cite{Kumar2015}, who found that a combination of host galaxy apparent magnitude and variability amplitude can be used to discriminate between AGN and SNe, where AGN have brighter host magnitudes and smaller variability amplitudes relative to their host galaxy flux.

\begin{figure}
\centering
\includegraphics[width=3.5in, angle=0]{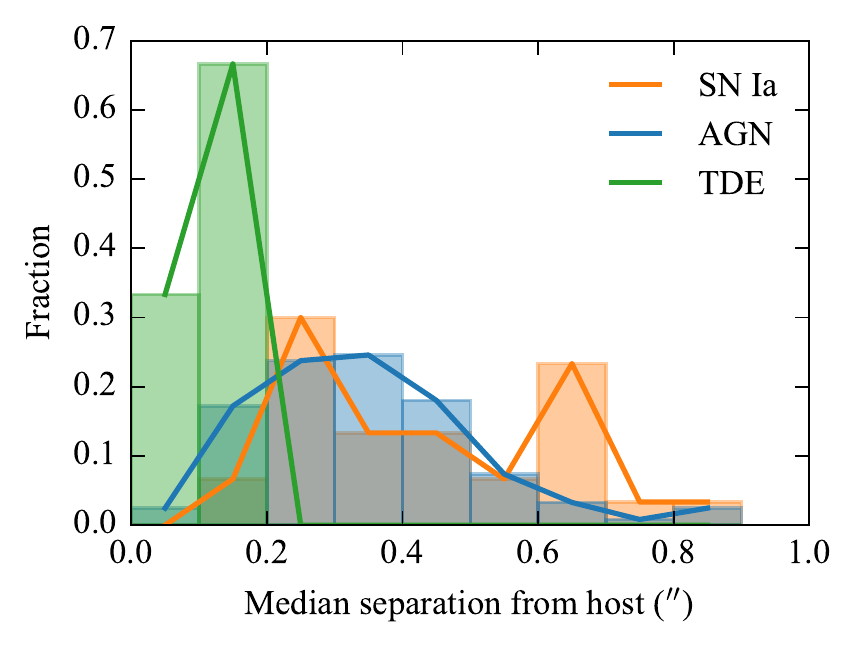}
\caption{Distribution of the median flare-host separation in the first week of discovery for AGN, SN Ia, and TDEs. Another 41\% of the SNe Ia can be removed from our nuclear sample if we employ a spatial cut at $d<$ 0.5\arcsec{}.}
\label{fig:sep}
\end{figure}

In \autoref{subsec:step1(nuclear)} we removed flares that are farther than 0.8\arcsec{} away from the centroid of their host galaxies. This offset was calculated by taking the median of all offsets measured from each detection up to the time when the transient was saved. We investigate this condition further by plotting the distributions of the flare-host separation for our control sample with spectroscopically confirmed AGNs (34), SNe Ia (26), and TDEs (3). The histograms of the median separation in the first week of discovery are plotted in \autoref{fig:sep}.

We noticed that the flare-host separation for AGN peaks toward 0.3\arcsec{} while the separation for SN Ia peaks around 0.2\arcsec{} and 0.6\arcsec{}. The TDEs, with a small sample size of 3, all seem to have a nuclear offset of less than 0.2\arcsec{}. While it is unclear why there seem to be a double-peaked offset for SN Ia, the second peak of SN Ia clearly stands out from the distribution of the nuclear flares. If we place a tighter cut at $d<$ 0.5\arcsec{} we can get rid of 4 out of 14 (29\%) of the SNe Ia with the new spatial cut alone.

We compare the composition of our sample selected with the initial selection criteria and the empirical cuts defined in this section in \autoref{fig:pie}. With the initial selection, we end up with a TDE contamination rate of 13:1 (left panel in \autoref{fig:pie}). By requiring \gptf{}$-$\rptf{}$<-$0.2 mag to remove AGNs and SNe, we achieve a contamination rate of 6:1 (middle panel in \autoref{fig:pie}). After applying the spatial cut at $d<$ 0.5\arcsec{} and host galaxy magnitude cut $r_\textrm{PSF} < $21.0 mag, our lowest TDE contamination rate is 4.5:1 (right panel in \autoref{fig:pie}).

\subsection{TDE Rate}
\label{subsec:TDErate}
The number of TDEs detected by the survey can be expressed as

\begin{equation}
\label{eq:rate}
N_{TDE} = \int \frac{d\dot{N}}{dL_g} \times \frac{4\pi}{3}D_{max}^3(L_g) \times \frac{\Delta \Omega}{4\pi} \times \tau~dL_g
\end{equation}
where $d\dot{N}/{dL_g}$ is the luminosity function of TDEs (the volumetric TDE rate with respect to peak \gptf{} band luminosity
), $D_{max}$ is the maximum distance (redshift) within which a flare can be detected by the flux-limited survey and is a function of the peak magnitude of the flare, $\Delta \Omega$ is the survey area, and $\tau$ is the duration of the survey.

We estimate the average timescale of the survey over all the fields by

\begin{equation}
\tau = \sum_{i=1}^{N} \Delta \tau_{max} / N,
\end{equation}
where i is the i-th field observed by iPTF, $\Delta \tau_{max}$ is the longest baseline in the \gptf{}+\rptf{} survey for the i-th field (right panel of \autoref{fig:grobs}), and N is the total number of unique fields. For fields that are observed before and after the hiatus in summer, we use the longest $\tau_{max}$ of these two periods to represent the baseline for the field. By summing over all the unique fields, we find $\tau\sim$44 days. Since the tidal disruption flares are luminous and long-lasting, the detection is not be sensitive to the week-long observing gaps caused by the full moon. This is true even for fast events like iPTF16fnl, which peaked at \gptf{}$=$17.07 mag and faded below 20th mag after 70 days. For iPTF16fnl-like events, which fade by 0.4 mag in the first 7 days post peak, they should remain detectable by iPTF out to $z\sim$~0.05 even in the presence of a week-long gap in the light curve.

We can solve for $D_{max}$ by requiring the apparent magnitude of a source with a given peak luminosity $\nu L_g$ being above the limiting magnitude $m_g$ = 20 mag. We assume a standard blackbody temperature of $T$ = 3$\times$10$^4$ K, which is representative for the optical TDEs, to account for K-correction. The differential flux is expressed as
\begin{equation}
    \label{eq:diff_flux}
    f_\nu = \frac{(1+z)L_{\nu(1+z)}}{4\pi D_L^2(z)},
\end{equation}
where $f_\nu$ is the observed flux density, $L_{\nu(1+z)}$ is the emitted luminosity density, and $D_L$ is the luminosity distance of the object. We then solve $z$ by converting $f_\nu$ back to AB magnitude.

The empirical luminosity function of UV/optical selected TDEs is measured in \cite{2017arXiv170703458V}, which can be approximated as a steep power-law where $d\dot{N}/{dL_g}$ $\propto$ $L_g^{-2.5}$. The luminosity function can then be parameterized at $L_0$ = 10$^{43}$ erg s$^{-1}$ as

\begin{equation}
    \mathrm{log_{10}} \left (\frac{d\dot{N}}{d \mathrm{log_{10}}L_g} \right)= \mathrm{log_{10}}(\dot{N_0}) - 1.5 \mathrm{log_{10}}{\left(\frac{L_g}{L_0}\right)}
\end{equation}

We adopt the power-law form of luminosity function from \cite{2017arXiv170703458V} and substitute in the areal coverage (4792 deg$^2$) and timescale ($\tau=$44 days) into \autoref{eq:rate}. We integrate \autoref{eq:rate} from 10$^{42}$ -- 10$^{43.5}$ erg s$^{-1}$. The lower integration limit is chosen to include iPTF16fnl, which has an observed peak luminosity of 10$^{42.3}$ erg s$^{-1}$ in $g$ band. The upper integration limit is chosen since the luminosity bin is not well-sampled beyond $\nu L_g > $ 10$^{43.5}$ erg s$^{-1}$. The only TDE in this high luminosity bin is ASASSN-15lh, which is thought to be a TDE around a rotating black hole with a mass $>$ 10$^{8}$ \Msun{} \citep{2016NatAs...1E...2L}. We also note that the nature of this event is still debated as it is also interpreted as a highly super-luminous supernova in \cite{2016Sci...351..257D}. Our chosen luminosity range would also include iPTF16axa, which has an observed peak luminosity $\nu L_g$ of 10$^{43}$ erg s$^{-1}$. Low luminosity TDEs like iPTF16fnl should be harder to see in flux-limited surveys since its detection volume is almost two orders of magnitude smaller than that for events like iPTF16axa. The fact we are seeing both events may suggest a steep TDE luminosity function.

We estimate the pipeline efficiency from \cite{2017ApJS..230....4F}, where they investigated the efficiency as a function of the ratio of host galaxy surface brightness to the flux of the flare. Since they estimated the host surface brightness by integrating the counts over a small image size that is close to the size of the PSF, we use the PSF magnitude from PS1 as a proxy for the surface brightness in our calculation. Our variability amplitude ($\Delta m_{var}$) cut already places a limit on this ratio. We are only sensitive to flares with $F_{\textrm{PSF, host}}/F_{\textrm{PSF, flare}}$ $\lesssim$ 1.6 since we required a $\gtrsim$ 60\% increase in the nuclear region in step 5. The median of the $F_{\textrm{PSF, host}}/F_{\textrm{PSF, flare}}$ ratio is $\sim$ 0.6 in our sample after applying the $\Delta m_{var}$ cut. This value corresponds to the $\approx$50\% recovery fraction \citep[][Fig 6]{2017ApJS..230....4F}.

After applying an efficiency factor of $\epsilon$ = 0.5, the 2 TDEs discovered in the iPTF color experiment would imply a rate of $\dot{N_0}$~=~1.1 $^{+1.8}_{-0.8}$ $\times$ 10$^{-7}$ Mpc$^{-3}$ yr$^{-1}$. We note that the volumetric rate is a lower limit because our selection is only sensitive to the red galaxies. We account for this bias when calculating the per galaxy rate as detailed below.

The volumetric rate can be converted to a per galaxy rate $\dot{n}$ by dividing the galaxy density that can be estimated from the galaxy luminosity function. We use the luminosity function derived for SDSS in \cite{2001AJ....121.2358B}, which has the form of a Schechter function, to calculate galaxy density ($\rho$) probed by iPTF. The majority (90\%) of the red host galaxies with spectroscopic redshift in our sample have $r$ band magnitudes falling in the range $-$22.7$<M_r<$$-$18.9. Since the more massive black holes cannot produce a tidal disruption flare, we cut off the luminosity at $M_r$ = $-$21.5 mag, which corresponds to a black hole mass of 10$^8$ \Msun{} in \cite{2007ApJ...663...53T}.

\begin{figure}
\centering
\includegraphics[width=3.5in, angle=0]{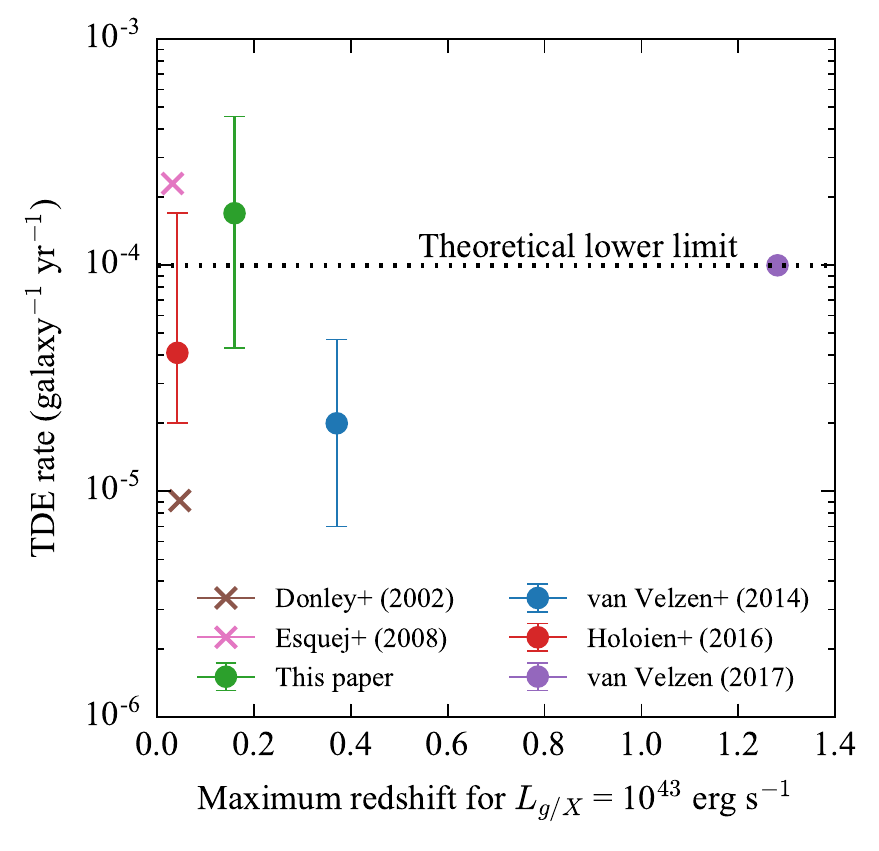}
\caption{Observational rate of TDEs from X-ray (crosses) and optical (circles) surveys including SDSS+PS1 \citep{2014ApJ...792...53V}, ASASSN \citep{Holoien2016}, iPTF (this paper), ROSAT All-Sky
Survey \citep{2002AJ....124.1308D}, and XMM-Newton Slew Survey \citep{2008A&A...489..543E}. The x-axis shows the maximum redshift for each survey to detect a flare with a peak luminosity of $L_g$ = 10$^{43}$ erg s$^{-1}$ for optical surveys or $L_X$ = 10$^{43}$ erg s$^{-1}$ in 0.2--2.4keV energy band for X-ray surveys. The dotted horizontal line marks the theoretical lower limit from \cite{2004ApJ...600..149W}.}
\label{fig:rate}
\end{figure}

\begin{figure}
\centering
\includegraphics[width=3.5in, angle=0]{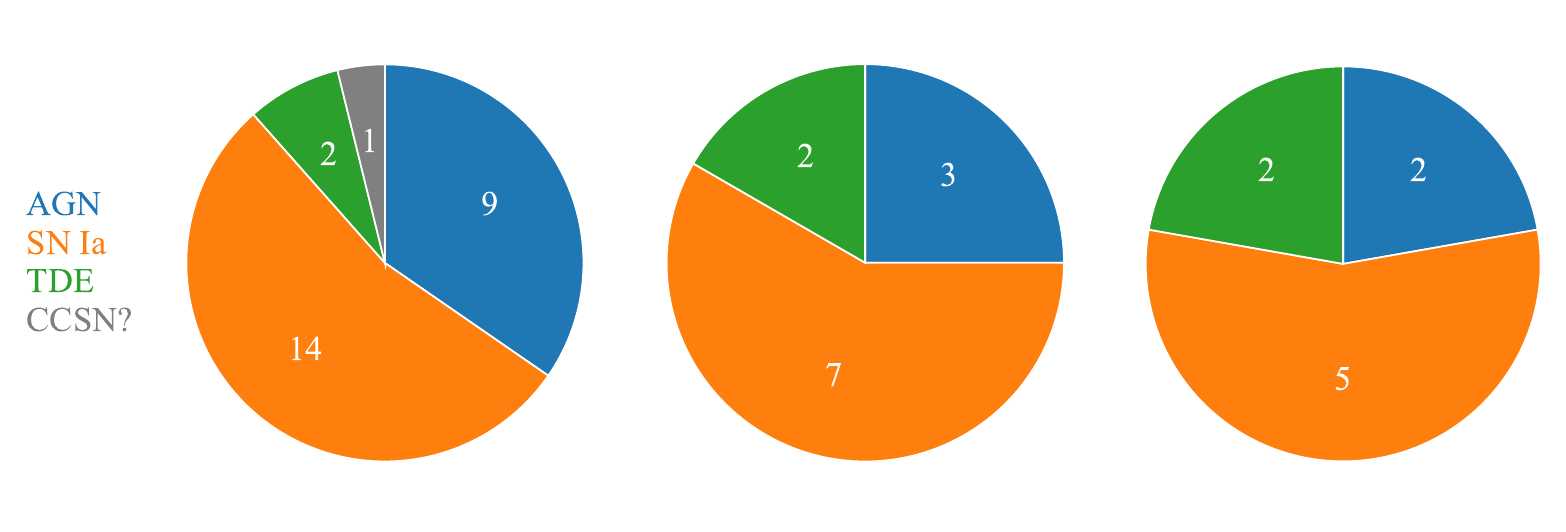}
\caption{$Left$: Pie charts of the classification in the final sample of 26 sources. $Middle$: After applying the color cut \gptf{}$-$\rptf{} $<$ $-$0.2, 6 AGNs and 5 SNe Ia are removed from the sample . $Right$: After applying the spatial cut $d<$ 0.5\arcsec{} and the host galaxy magnitude cut $r_\textrm{PSF} < $21.0 mag, we are left with 2 TDEs, 5 SNe Ia, and 2 AGNs.}
\label{fig:pie}
\end{figure}

We also account for the fraction of red galaxies ($u-g >$ 1.0 mag and $g-r>$ 0.5 mag) in the luminosity range of $-$21.5$<M_r<$$-$18.9 by querying the SDSS database. This fraction is 21\% for sources within SDSS footprint ($ugr$) and 74\% for sources with PS1 photometry ($gr$). Using the source number weighted fraction of 34.8\%, we estimate a density $\rho$ of 6.4$\times$10$^{-4}$ gal Mpc$^{-3}$, implying a TDE rate of $\approx$ 1.7 $^{+2.9}_{-1.3}$ $\times$10$^{-4}$ TDEs gal$^{-1}$ yr$^{-1}$. We compare the per galaxy rate derived in this study to the value reported from other X-ray and optical surveys \autoref{fig:rate}.

Lastly, we note that the TDE rate derived here only applies to TDEs characteristic of the known optically-selected TDE population in inactive galaxies. Our selection method of excluding AGNs and blue galaxy hosts may have the caveat of introducing a systematic error in the TDE rate since the types of TDEs associated with these galaxies are not probed in this work.

\subsection{Prospects with ZTF}

The camera mounted on P48 has been replaced with a wider FOV camera (47 deg$^2$) for ZTF, which is almost an order of magnitude upgrade from iPTF. Together with the reduced overhead time, ZTF will be scanning the sky at a rate of 3760 deg$^2$ hr$^{-1}$. In 2018, 40\% of the ZTF operation time will be dedicated to the public survey to monitor $\sim$15000 deg$^2$ of the sky with a 3-day cadence with near-simultaneous \gptf{} and \rptf{} observations.

We define a field as visible at Palomar to have an airmass$<$3 for more than 3 hours in one night. The baseline of each field is $\sim$8 months. Scaling from this iPTF color experiment to the wider areal coverage and longer baseline for ZTF, we expect to find $\sim$ 2$\cdot (\frac{15000}{4792})(\frac{8 \mathrm{months}}{44 \mathrm{days}})$ = 32$^{+41}_{-25}$ TDEs in a year.

We estimate ZTF to discover $(\frac{15000 \textrm{deg}^2/3\textrm{day}}{771 \textrm{deg}^2})$ $\times$ 6 nuclear candidates per night = 39 nuclear candidates per night. The original selection criteria in our nuclear transient pipeline can already get rid of $\sim$ 95\% of the transients, reducing the number of candidates to 2 per night.

We show the classification of the final sample in a pie chart in \autoref{fig:pie}. By employing the empirical cuts (\gptf{}$-$\rptf{} $<$ $-$0.2 mag, $d<$0.5\arcsec{}, $r_\textrm{PSF, host} < $21.0 mag) established in this paper, we are able to further reduce the number of contamination from 13:1 down to 4.5:1 (right panel in \autoref{fig:pie}). The number of nuclear transients satisfying these new criteria will be 0.7 per night for ZTF.

During the operation of ZTF, the low resolution SEDm will be entirely dedicated to spectroscopic follow up of ZTF transients that are brighter than 19 mag in \gptf{}.
In our sample, $\sim$20\% of the sources have apparent magnitudes $\lesssim$ 19.0 mag near peak, where rapid SEDm classification will be feasible. $u$ band imaging and higher precision astrometry will also help to distinguish between SNe Ia and TDEs, leaving only the highest-confidence TDE candidates. With our selection criteria, the number of candidates that require follow up will be a manageable amount for available spectroscopic resources. With a spectroscopically complete sample of TDEs, we will be able to measure TDE luminosity function and rate that are closely connected to the physical processes leading to the disruption of stars by the central black holes of their host galaxies.

\section{Acknowledgements}
We thank the anonymous referee for helpful comments that improved the manuscript. T.H. thanks Jesper Sollerman for his feedback on the manuscript. S.G. is supported in part by NASA Swift Cycle 12 grant NNX16AN85G and NSF CAREER grant 1454816.  These results made use of the Discovery Channel Telescope at Lowell Observatory. Lowell is a private, non-profit institution
dedicated to astrophysical research and public appreciation of astronomy and operates the DCT in partnership
with Boston University, the University of Maryland, the
University of Toledo, Northern Arizona University, and
Yale University. The W. M. Keck Observatory is operated as a scientific partnership among the California
Institute of Technology, the University of California, and
NASA; the Observatory was made possible by the generous financial support of the W. M. Keck Foundation.
This research used resources of the National Energy Research Scientific Computing Center, a DOE Office of Science User Facility supported by the Office of Science of
the U.S. Department of Energy under Contract No. DEAC02-05CH11231.

\begin{deluxetable*}{lccccccc}
\tablecaption{Photometric properties of the final sample \label{tab:sample}}
\tablecolumns{8}
\centering
\tablehead{\colhead{Object} & \colhead{redshift} & \colhead{$m_{peak,g}$} & \colhead{$M_{peak,g}$} & \colhead{$\Delta m_{var}$} & \colhead{g-r}& \colhead{Classification} & \colhead{Instrument$^1$}}
\startdata
iPTF16axa$^{s}$ & 0.108 & 19.44 & -19.08 & -1.16 & -0.45 & TDE & DEIMOS \\
iPTF16ayd & 0.171 & 19.54 & -20.06 & -0.54 & -0.01 & AGN & LRIS \\
iPTF16bcp$^{s}$ & 0.14$^{f}$ & 19.55 & -19.57 & -3.10 & -0.19 & SN Ia & DEIMOS \\
iPTF16bjy & 0.128 & 19.65 & -19.27 & -0.52 & -0.20 & SN Ia phot & \\
iPTF16bke & 0.15$^{f}$ & 19.25 & -20.01 & -0.72 & -0.28 & SN Ia & LRIS \\
iPTF16bko & 0.085 & 18.74 & -19.22 & -1.09 & -0.10 & SN Ia phot & \\
iPTF16bmy & 0.129 & 19.94 & -18.99 & -1.04 & -0.14 & CCSN? & \\
iPTF16fhs & 0.110 & 19.08 & -19.48 & -0.81 & -0.12 & AGN phot &  \\
iPTF16fjc & 0.449 & 19.56 & -22.45 & -2.01 & -0.11 & AGN phot &  \\
iPTF16fjg & 0.14$^{f}$ & 19.3 & -19.83 & -0.88 & -0.23 & SN Ia & SEDM \\
iPTF16fly & 0.349 & 18.38 & -22.99 & -1.25 & -0.24 & AGN phot &  \\
iPTF16fmc & 0.139 & 19.37 & -19.75 & -1.09 & -0.34 & SN Ia phot & \\
iPTF16fmd$^{s}$ & 0.151 & 19.75 & -20.05 & -1.86 & -0.17 & SN Ia phot & \\
iPTF16fnl$^{s}$ & 0.016 & 17.06 & -17.21 & -0.78 & -0.22 & TDE & SEDM \\
iPTF16fpt & 0.160 & 19.52 & -19.93 & -0.75 & -0.12 & AGN phot &  \\
iPTF16fqa$^{s}$ & 0.125 & 19.1 & -19.76 & -0.51 & -0.21 & AGN & SEDM \\
iPTF16fsc & 0.167 & 19.23 & -20.30 & -0.64 & -0.02 &  SN Ia phot &  \\
iPTF16fzx & 0.125 & 19.73 & -19.13 & -0.60 & -0.10 & AGN phot &  \\
iPTF16glz & 0.03$^{f}$ & 16.91 & -18.7 & -5.16 & -0.25 & SN Ia & SEDM \\
iPTF16gyl$^{s}$ & 0.112 & 19.82 & -18.78 & -0.79 & -0.08 & SN Ia phot & \\
iPTF16hcn & 0.130 & 19.59 & -19.36 & -0.99 & -0.28 & SN Ia phot &  \\
iPTF16hma$^{s}$ & 0.391 & 19.48 & -22.18 & -1.35 & -0.20 & AGN phot &  \\
iPTF16hrx & 0.105 & 19.29 & -19.16 & -0.51 & -0.28 & SN Ia phot & \\
iPTF16ijw & 0.07$^{f}$ & 18.15 & -19.36 & -1.25 & -0.20 & SN Ia & SEDM \\
iPTF16ijz & 0.151 & 19.56 & -19.73 & -0.71 & -0.23 & AGN phot &  \\
iPTF16isi & 0.092 & 19.45 & -18.67 & -0.75 & -0.30 & SN Ia phot &
\enddata
\tablenotetext{1}{Objects for which no classification spectrum was taken are left blank.}
\tablenotetext{s}{Has \Swift{} $uvw2$ photometry. Results are summarized in \autoref{tab:swiftsample}.}
\tablenotetext{f}{Redshift derived from SNID fit to the flaring SN Ia spectra.}
\end{deluxetable*}

\begin{deluxetable}{lcccccc}
\tablecaption{Best-fit parameters to the SALT2 model \label{tab:SALT2_params}}
\tablecolumns{7}
\centering
\tablehead{\colhead{Object} & \colhead{$\chi^2_\nu$} & \colhead{redshift} & \colhead{$t_0$} & \colhead{$x_0$ ($\times$10$^{-4}$)} &\colhead{$x_1$}& \colhead{c}}
\startdata
iPTF16bcp & 1.03 & 0.14 & 57545.6 & 2.47 & 1.86 & -0.137 \\
iPTF16bke & 4.62 & 0.15 & 57558.4 & 2.57 & 1.87 & -0.162 \\
iPTF16fjg & 1.54 & 0.14 & 57627.4 & 3.2 & 0.83 & -0.217 \\
iPTF16glz & 2.13 & 0.03 & 57664.8 & 36.67 & 0.98 & -0.024 \\
iPTF16ijw & 1.73 & 0.07 & 57722.1 & 10.55 & 2.66 & -0.005 \\
iPTF16bjy & 1.70 & 0.128 & 57550.1 & 2.0 & 3.51 & -0.14 \\
iPTF16bko & 0.49 & 0.085 & 57549.1 & 6.02 & 1.29 & -0.086 \\
iPTF16fmc & 0.89 & 0.139 & 57639.4 & 3.15 & 2.52 & -0.108 \\
iPTF16fmd & 1.05 & 0.151 & 57637.7 & 2.26 & 1.18 & -0.082 \\
iPTF16fsc & 2.07 & 0.167 & 57641.4 & 1.31 & 6.18 & 0.021 \\
iPTF16gyl & 2.48 & 0.112 & 57672.9 & 1.93 & 0.37 & -0.006 \\
iPTF16hcn & 5.18 & 0.130 & 57670.6 & 2.64 & -0.17 & -0.096 \\
iPTF16hrx & 1.62 & 0.105 & 57699.5 & 2.99 & -1.04 & -0.077 \\
iPTF16isi & 1.88 & 0.092 & 57725.5 & 2.8 & 1.88 & -0.185
\enddata
\tablenotetext{1}{Best-fit redshift from SALT2 model.}

\end{deluxetable}

\begin{deluxetable}{lccccc}
\tablecaption{Observations of the host galaxies of the photometrically classified sources \label{tab:host_obs}}
\tablecolumns{6}
\centering
\tablehead{\colhead{Object} & \colhead{$z$} & \colhead{Date} & \colhead{Telescope} & \colhead{Host class$^{1}$} & \colhead{FWHM$_{H\beta}$ (km s$^{-1}$)}}
\startdata
iPTF16bjy	&0.128	&2017-09-15	&DCT Deveny & broad line AGN  & 6199.0\\
iPTF16bko	&0.085	&2017-09-15	&DCT Deveny & early-type galaxy & \nodata\\
iPTF16bmy	&0.129	&2017-07-29	&P200 DBSP &  early-type galaxy & \nodata\\
iPTF16fhs	&0.110	&2017-09-15	&DCT Deveny & AGN & \nodata\\
iPTF16fjc	&0.449	&2017-09-17	&DCT Deveny & broad line AGN & 4086.3\\
iPTF16fly	&0.349	&2017-09-15	&DCT Deveny & broad line AGN & 4633.2\\
iPTF16fmc	&0.139	&2017-07-29	&P200 DBSP & starforming & \nodata\\
iPTF16fmd	&0.151	&2017-07-29	&P200 DBSP & AGN & \nodata \\
iPTF16fpt   &0.160	&2012-06-29 &SDSS(host) & AGN & \nodata \\
iPTF16fzx	&0.125	&2017-09-15	&DCT Deveny & starforming & \nodata\\
iPTF16gyl	&0.112	&2017-09-17	&DCT Deveny & AGN & \nodata\\
iPTF16hcn	&0.130	&2017-09-16	&DCT Deveny & AGN & \nodata\\
iPTF16hma	&0.391	&2017-07-31	&P200 DBSP & broad line AGN & 4419.3\\
iPTF16hrx	&0.105	&2017-09-15	&DCT Deveny & starforming & \nodata\\
iPTF16ijz	&0.151	&2017-09-15	&DCT Deveny & AGN & \nodata\\
iPTF16isi	&0.092	&2017-09-15	&DCT Deveny & starforming & \nodata
\enddata
\tablenotetext{1}{Host galaxies are classified based on the [OIII]/H$\beta$ versus [NII]/H$\alpha$ diagnostic diagram.}
\end{deluxetable}

\begin{deluxetable*}{llcclccccc}{10}
\tablecaption{\Swift{} observations of the sample \label{tab:swiftsample}}
\centering
\tablehead{\colhead{Object} &\colhead{Phase} & \colhead{UVW2$^1$} & \colhead{GALEX NUV} & \colhead{XRT} & \colhead{$z$}& \colhead{$m_{peak,g}$} & \colhead{$m_{peak,r}$} & \colhead{Class}
& \colhead{$\Delta m_{var}$} \\
\colhead{} & \colhead{days} & \colhead{mag} & \colhead{mag} & \colhead{erg s$^{-1}$cm$^{-2}$}& \colhead{} & \colhead{}  & \colhead{} & \colhead{}}
\startdata

iPTF16axa   & 8.7$^d$ &  18.77 $\pm$0.05   & \nodata     & \textless1$\times$10$^{-14}$  & 0.108 & 19.44 & 19.99 & TDE    & -1.16   \\
iPTF16bcp   & 1.7$^d$ &  \textgreater22.60 & \nodata     & \textless3.3$\times$10$^{-13}$  & 0.14& 19.55 & 19.77 & SN Ia  & -3.10   \\
iPTF16fmd   & 3.8 &  \textgreater22.02 & \nodata     & \textless4.1$\times$10$^{-13}$  & 0.151   & 19.75 & 19.62 & SN Ia phot   & -1.86   \\
iPTF16fnl   & 0.6 &  16.32 $\pm$0.03   & 19.57  &  4.6$^{+3.6}_{-2.0}$$\times$10$^{-15}$ & 0.016 & 17.06 & 17.23 & TDE    &  -0.78  \\
iPTF16fqa   & 13.8$^d$ &  19.23 $\pm$0.06   & 20.59 & \textless2.7$\times$10$^{-13}$  & 0.125 & 18.77 & 18.87 & AGN    & -0.51 \\
iPTF16gyl   & 8.5 &  22.03 $\pm$0.31   & 22.81  & \textless2.7$\times$10$^{-13}$  & 0.112  & 19.82 & 19.35 & SN Ia phot   & -0.79 \\
iPTF16hma   & 69.5$^d$ &  19.78 $\pm$0.08   & \nodata     & \textless5.9$\times$10$^{-13}$  & 0.391 & 19.47 & 19.65 & AGN phot   & -1.35

\enddata
\tablenotetext{1}{Galactic extinction correction applied.}
\tablenotetext{d}{If the peak is not clearly sampled in the light curve, we list $\Delta t$ since iPTF discovery instead.}
\end{deluxetable*}
\clearpage
\bibliography{tde}

\appendix

\renewcommand{\thefigure}{A\arabic{figure}}
\setcounter{figure}{0}

\begin{figure*}
\centering
\includegraphics[width=7.0in, angle=0]{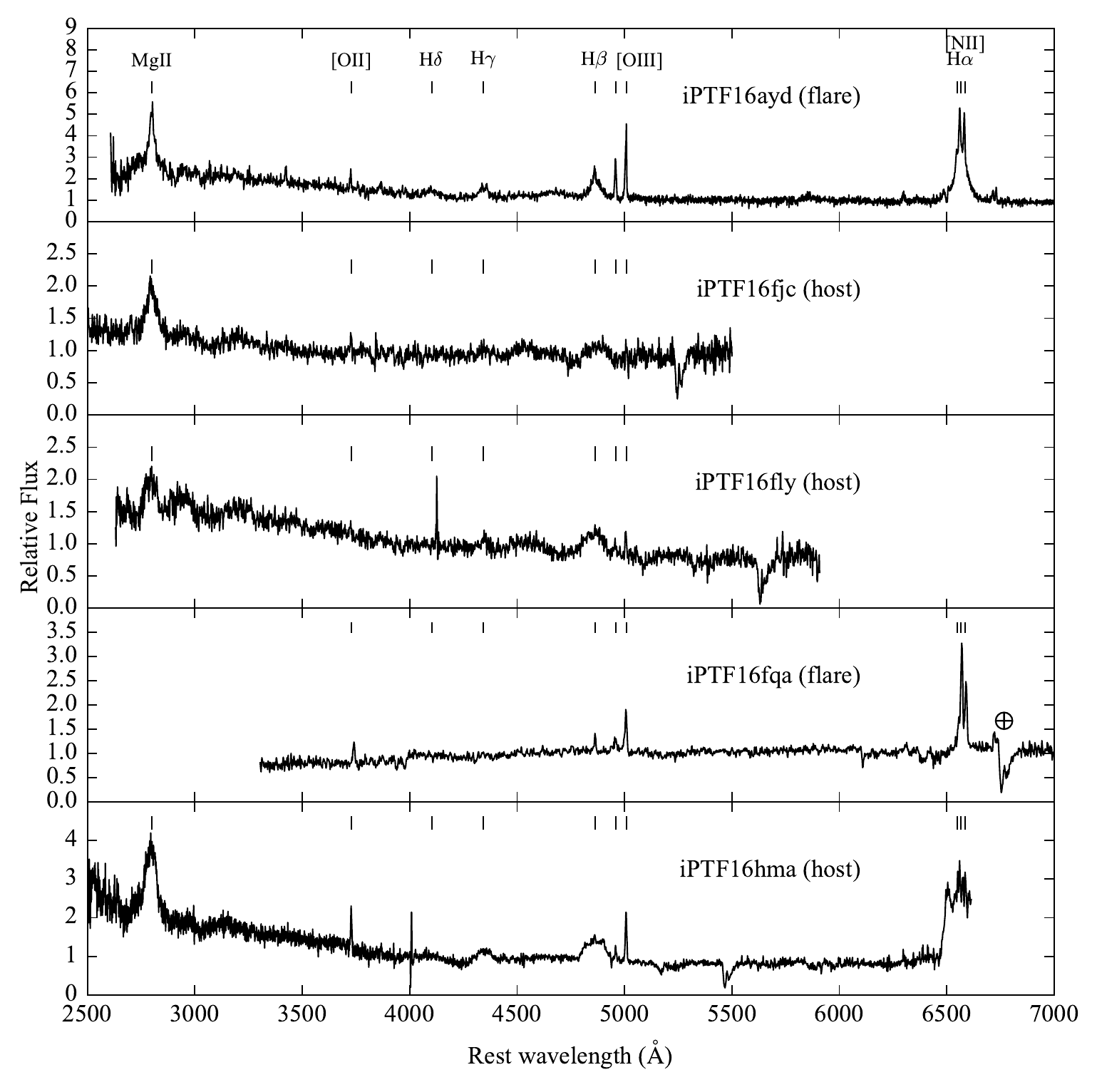}
\caption{AGNs classified based on the presence of broad emission lines. Strong telluric absorptions are marked with the $\oplus$ symbol.}
\label{fig:broadlineAGN}
\end{figure*}

\begin{figure*}
\centering
\includegraphics[width=7.0in, angle=0]{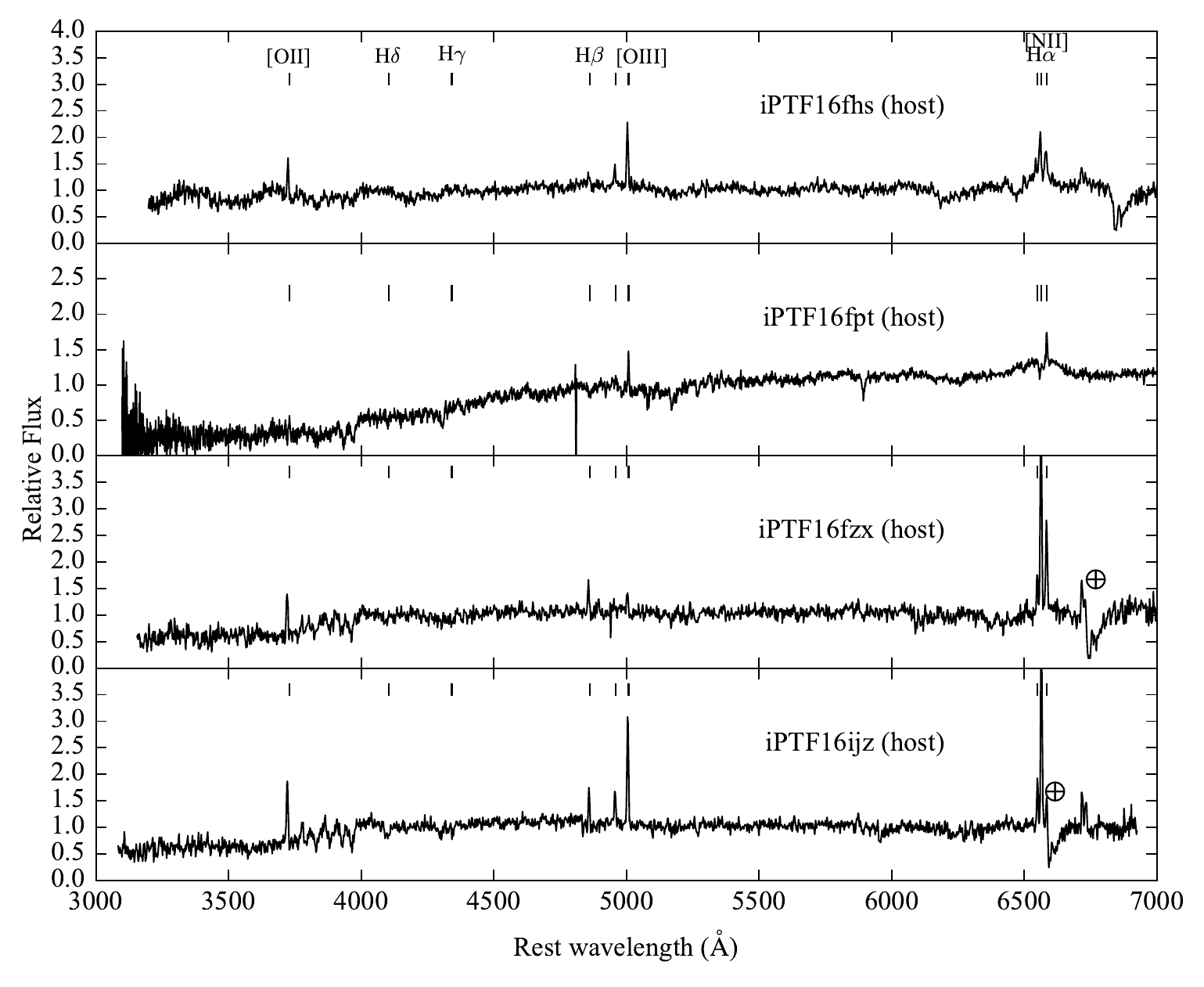}
\caption{AGNs classified by [OIII]/H$\beta$ versus [NII]/H$\alpha$ line ratio. Strong telluric absorptions are marked with the $\oplus$ symbol.}
\label{fig:type2AGN}
\end{figure*}

\begin{figure*}
\centering
\includegraphics[width=7.0in, angle=0]{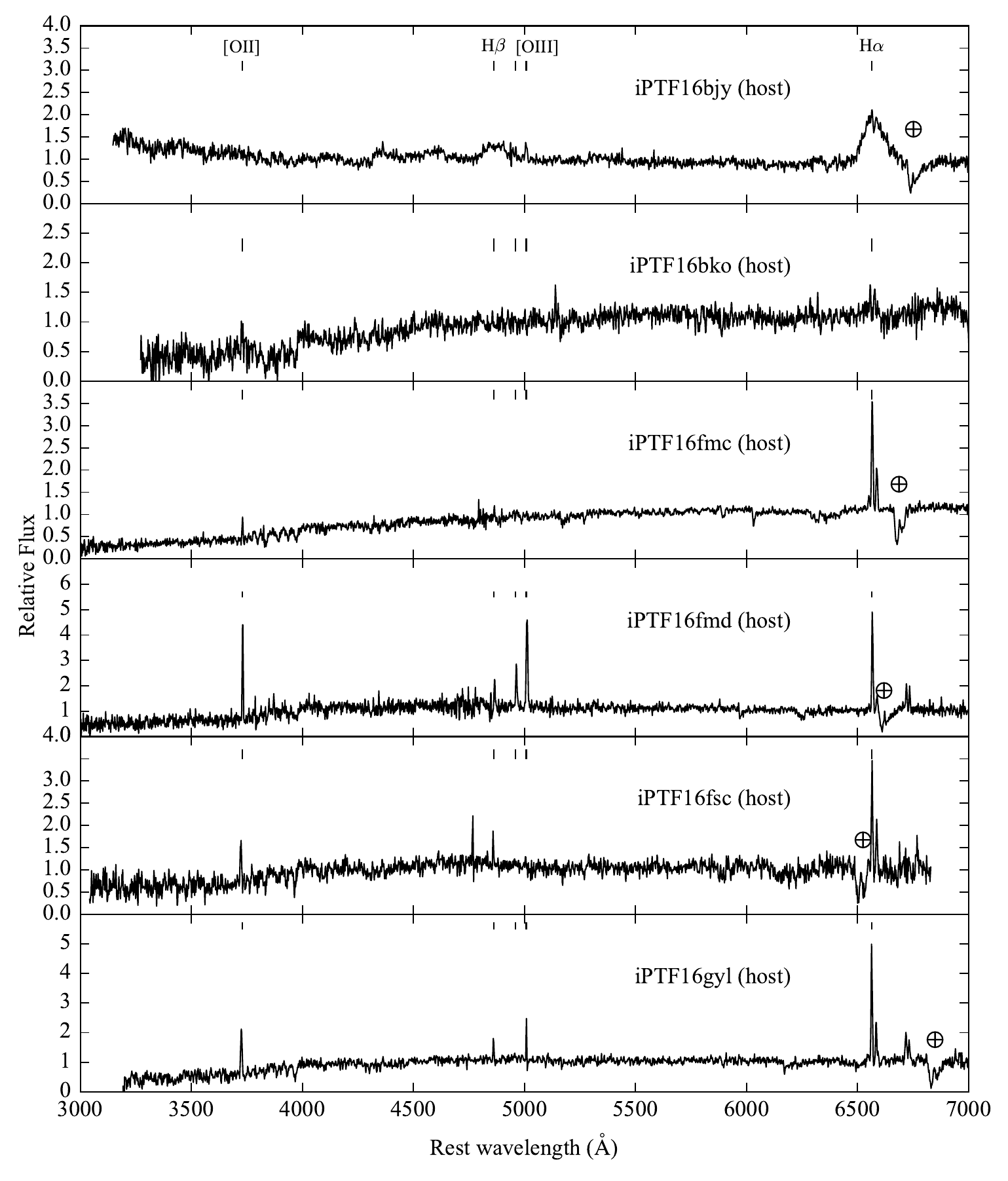}
\caption{Host galaxies of SNe Ia classified based on the photometry. Strong telluric absorptions are marked with the $\oplus$ symbol.}
\label{fig:SNI}
\end{figure*}

\addtocounter{figure}{-1}

\begin{figure*}
\centering
\includegraphics[width=7.0in, angle=0]{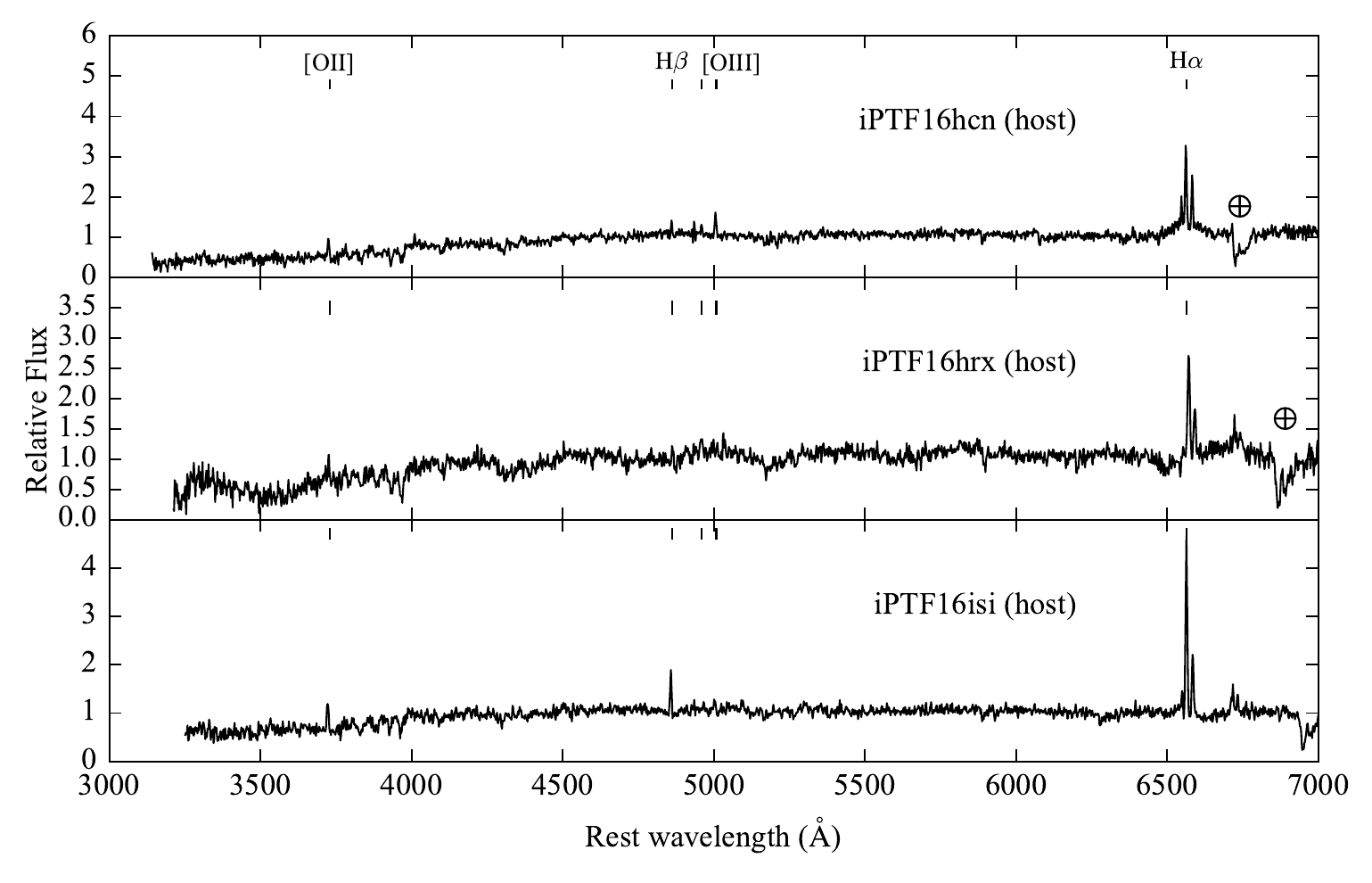}
\caption{Continued.}
\end{figure*}

\begin{figure*}
\centering
\includegraphics[width=7.0in, angle=0]{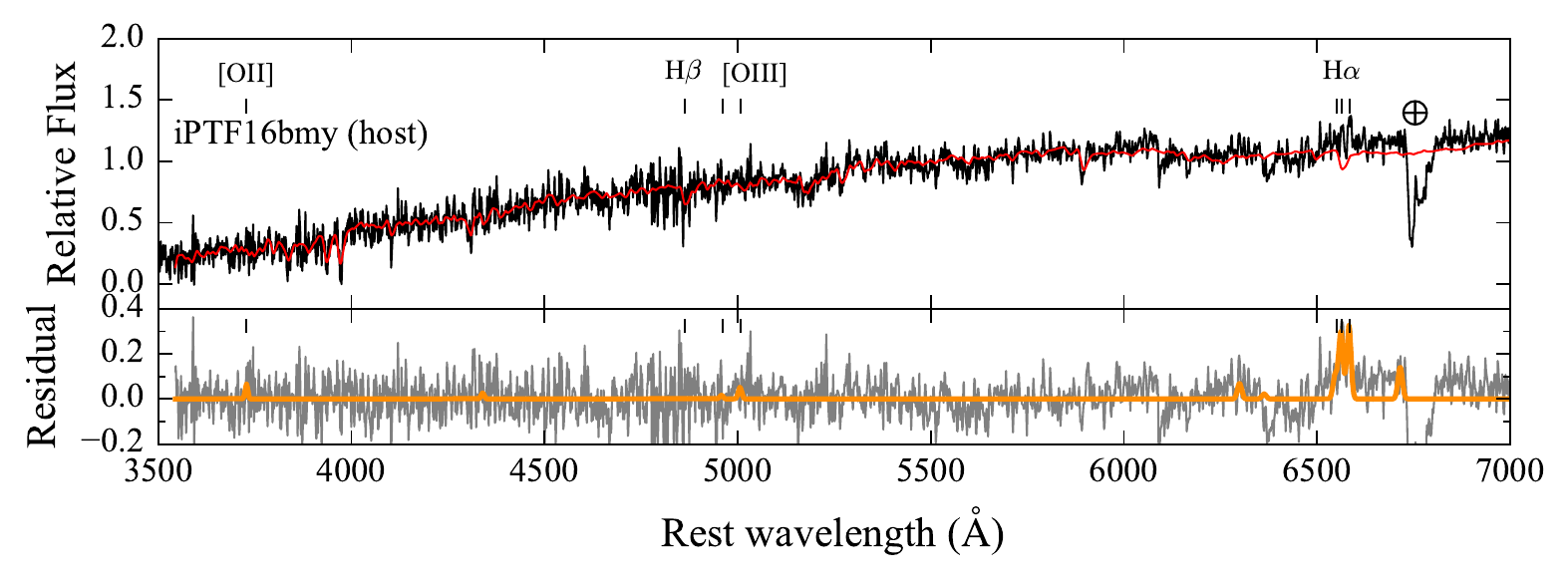}
\caption{Host spectrum of iPTF16bmy. The black line in the top panel shows the P200 spectrum of the host galaxy of iPTF16bmy while the red line shows the best-fit galaxy spectrum from \texttt{ppxf}. The bottom panel shows the residual of the fit in grey. The emission line profiles that were simultaneously fitted with the stellar templates are colored in orange. Weak H$\alpha$, [NII], and [OIII] lines detected in the host spectrum are suggestive of ongoing starformation.}
\label{fig:bmyhost}
\end{figure*}

\end{document}